\documentclass[a4paper,11pt]{article}
\pdfoutput=1
\usepackage{jheppub}
\usepackage{amsmath}
\usepackage{amssymb}
\usepackage{color}
\usepackage[dvipsnames]{xcolor}
\usepackage{graphicx}
\usepackage{hyperref}
\usepackage{xspace,slashed}
\usepackage[utf8]{inputenc}
\usepackage{subcaption}
\usepackage{multirow}
\usepackage{nccmath} 
\usepackage{orcidlink}
\usepackage{subcaption}
\usepackage{cleveref}
\usepackage{tikz}
\usepackage[compat=1.1.0]{tikz-feynman}
\usepackage{comment}
\usetikzlibrary{decorations.markings}
\usetikzlibrary{decorations.pathmorphing}
\usetikzlibrary{intersections}
\usetikzlibrary{calc}
\tikzset{
    vector/.style={
        decoration={snake, aspect=0.75, mirror, segment length=2mm},
        decorate
    },
    photon/.style={decorate, decoration={snake, amplitude=2pt, segment length=4pt}}
    ,
    gluon/.style={decorate, decoration={coil, amplitude=2pt, segment length=4pt}}
    ,
    fermion/.style={
    draw,
    thick,
    -{Latex[length=3mm,width=2mm]}
  }
}

\makeatletter
\renewcommand\@fpheader{}
\renewcommand\@journal{}
\makeatother

\newcommand{\ep}{\epsilon}
\newcommand{\oT}{{T}}
\newcommand{\oF}{{\mathcal{F}}}
\newcommand{\lambdavec}{{\vec{\lambda}}}
\DeclareMathOperator{\Li}{Li}

\preprint{}

\title{
Light-by-light scattering at three loops in massless QCD and QED: amplitudes and cross sections
}

\author[a]{Piotr Bargie\l{}a~\orcidlink{0000-0002-3646-5892}}
\emailAdd{pbargiel@ed.ac.uk}
\affiliation[a]{Higgs Centre for Theoretical Physics, School of Physics and Astronomy, The University of Edinburgh, Edinburgh EH9 3FD, Scotland, UK}

\author[b]{Amlan Chakraborty~\orcidlink{0000-0002-5952-4138}}
\emailAdd{amlan.chakraborty@unimi.it}
\affiliation[b]{Tif Lab, Dipartimento di Fisica, Universit\'{a} di Milano, Via Celoria 16, I-20133 Milano, Italy}

\author[c]{Giulio Gambuti~\orcidlink{0000-0003-4462-0423}}
\emailAdd{ggambuti@phys.ethz.ch}
\affiliation[c]{Institute for Theoretical Physics, ETH Z\"urich,
8093 Z\"urich, Switzerland}

\author[d]{Melih A. Ozcelik~\orcidlink{0000-0002-8312-7116}}
\emailAdd{melih.ozcelik@ijclab.in2p3.fr}
\affiliation[d]{Universit\'e Paris-Saclay, CNRS, IJCLab, 91405 Orsay, France}

\abstract{
We present the calculation of three-loop massless QCD and QED helicity amplitudes for light-by-light scattering. We make use of Lorentz tensor decomposition in the 't Hooft-Veltman dimensional regularisation scheme to reduce the complexity of the computation. 
Our analytic amplitude results are remarkably compact and can be efficiently evaluated numerically. We employ them to compute the corresponding NNLO differential cross-section predictions in the invariant mass and rapidity distributions of the di-photon system, for which we find agreement with the experimental ATLAS data from ultra-peripheral heavy-ion collisions.
}

\begin{document}

\maketitle

\section{Introduction}
\label{sec:intro}

Light is one of the most fundamental concepts in mankind's pursuit of studying Nature.
Its properties sparked the two major revolutions in modern Physics \textit{i.e.} the rise of Quantum Mechanics and Einstein's Relativity.
Classically, Maxwell's Electrodynamics equations forbid light self-interaction by the virtue of the superposition principle.
However, Quantum Electrodynamics (QED) admits light-by-light (LbL) scattering due to the interaction with virtual vacuum lepton-antilepton pairs.
As the three-photon process vanishes by Furry's theorem, the four-photon scattering considered in this work is the lowest-multiplicity example of light's nonlinearity.
LbL scattering has been investigated for around a century now (for a historical review see ref.~\cite{Scharnhorst:2017wzh}).
In this work, we provide the highest-precision theoretical prediction to date for this process in both massless QED and Quantum Chromodynamics (QCD).

LbL interactions can be experimentally observed at the Large Hadron Collider (LHC) in Ultra-Peripheral Collisions (UPCs) of heavy ions.
The first direct measurements have been recently performed by the ATLAS~\cite{ATLAS:2017fur,ATLAS:2019azn,ATLAS:2020hii} and CMS~\cite{CMS:2018erd,CMS:2024tfd} collaborations.
In UPC, the impact parameter of the colliding hadrons is rather large, hence the electromagnetic interaction dominates in the collision over the strong interaction.
As such, the cross section is enhanced by a large factor of $Z^4 \sim \mathcal{O}(10^7)$ arising from the two incoming heavy lead (Pb) ions with large charge number $Z = 82$ compared to the proton-proton case.
Following the newly available experimental LHC data, there has been a recent interest in stress testing LbL scattering against our understanding of the Standard Model (SM) of Particle Physics \textit{e.g.} by studying bound states of leptons~\cite{dEnterria:2022ysg,dEnterria:2023yao} and gluons~\cite{Greiner:1992fz}.
In addition, LbL serves as a probe for New Physics Beyond the SM (BSM) \textit{e.g.}
quartic anomalous gauge couplings~\cite{dEnterria:2013zqi}, 
graviton-like particles~\cite{dEnterria:2023npy,Atag:2010bh},
axion-like particles~\cite{Knapen:2016moh},  
large extra dimensions~\cite{Cheung:1999ja,Davoudiasl:1999di}, 
supersymmetry~\cite{Greiner:1992fz},
nonlinear Born-Infeld extensions of QED and SM~\cite{Ellis:2017edi}, 
and noncommutative QED~\cite{Horvat:2020ycy}.

There has been a lot of progress over the last decades in providing predictions for LbL scattering to higher perturbative orders.
In QED, the one-loop  amplitude and leading order (LO) cross section have been computed in ref.~\cite{Karplus:1950zz} for massless fermion loops.
For massive fermions, the amplitude was provided in ref.~\cite{Bernicot:2008th}, while the cross section is available in ref.~\cite{Bardin:2009gq}.
In electroweak theory, the one-loop amplitude mediated by $W$ bosons was calculated in ref.~\cite{Yang:1994nu}.
In QCD and QED the two-loop amplitude and the corresponding next-to-leading order (NLO) cross section were provided in ref.~\cite{Bern:2001dg} with massless fermion loops, while in ref.~\cite{AH:2023ewe} and ref.~\cite{AH:2023kor} for massive fermions, respectively.
In this paper, we compute the three-loop massless QCD and QED helicity amplitudes and the corresponding next-to-next-to-leading order (NNLO) cross section.
We point out that the three-loop amplitudes to all the remaining massless QCD processes have been calculated in refs.~\cite{Caola:2021rqz,Bargiela:2021wuy,Caola:2021izf,Caola:2022dfa,Bargiela:2022lxz}.

The rest of this paper is organised as follows.
In \cref{sec:hel}, we describe the three-loop QCD scattering amplitude properties and the calculation workflow.
In \cref{sec:qed}, we provide the details on abelianisation from QCD to QED amplitudes.
In \cref{sec:ren}, we briefly cover the QCD and QED renormalisation of our amplitudes.
In \cref{subsec:pheno}, we discuss our phenomenological differential cross section prediction and the comparison with experimental data.
We end with concluding remarks in \cref{sec:concl}.

\section{Helicity amplitudes}
\label{sec:hel}

We consider virtual massless $SU(N_c)$ gauge theory corrections up to three loops to the scattering process
\begin{equation}
\gamma (p_1) + \gamma(p_2)  \to \gamma(-p_3) + \gamma(-p_4)\,,
\label{eq:aaaa}
\end{equation}
with four-momenta $p_i^\mu$ of external photons and number of colours $N_c$ of the gauge group.
These momenta are all massless $p_1^2=p_2^2=p_3^2=p_4^2=0$ and they satisfy momentum conservation $p_1+p_2+p_3+p_4=0$.
We introduce the standard Mandelstam variables together with the corresponding momentum conservation equation
\begin{equation}
s = (p_1+ p_2)^2\,,~ t = (p_1+p_3)^2\,,~u = (p_2+p_3)^2 \,\,, 
\quad s+t+u = 0 \,.
\label{eq:physregion}
\end{equation}
In the physical scattering region, we have $s>0$, $t<0$, $u<0$.
It is also convenient to define a dimensionless ratio
\begin{equation}
    x = -\frac{t}{s},
\end{equation}
so that $0 < x < 1$ in the physical region.

The colour structure of this amplitude is particularly simple as the external particles are colourless.
Indeed, the amplitude is a polynomial in the Casimir invariants of $SU(N_c)$
\begin{equation}
\begin{split}
T^a_{ij}T^a_{jk} =&\, C_F\, \delta_{ik} \,,\qquad C_F=\frac{N_c^2-1}{2 N_c} \,,
\\
f^{acd}f^{bcd} =&\, C_A\, \delta^{a b} \,,\qquad C_A=N_c \,,
\end{split}
\end{equation}
with $T^{a}_{ij}$ the generators of the fundamental representation and $f^{abc}$ the structure constants of the $su(N_c)$ Lie algebra.
We normalise group generators such that
\begin{equation}
  {\rm Tr}[T^a T^b] = T_F\, \delta^{a b},~~~~ T_F = \frac{1}{2}.
\end{equation}
In QCD, we have that $N_c=3$, $C_A=3$ and $C_F=4/3$. In addition, the amplitude is a polynomial in the charge-dependent terms
\begin{equation}
    \tilde{n}_q^{(i)} = \sum_{n_q} \, e_q^i \,,
    \label{eq:nfdef}
\end{equation}
with $e_q$ the relative fractional charge of the quark running in the loop, summed over all active quark flavours $n_q$. We have that $e_q=2/3$ for up-type quarks and $e_q=-1/3$ for down-type quarks. The dependence on the $\tilde{n}_q^{(i)}$ terms arises from the multiple different ways in which the four photons can couple to massless closed fermion loops.
At this stage let us also define the following quantities 
\begin{equation}
    \alpha_b = \frac{e_b^2}{4\pi}\,, \quad
    \alpha_{s,b} = \frac{g_{s,b}^2}{4\pi}\, ,
    \label{eq:barecouplingdefQCDQED}
\end{equation}
where $e_b$ and $g_{s,b}$ are respectively the bare electric charge and the bare strong coupling. Our bare amplitudes will be expressed as series expansions in these parameters.\\

Let us now describe the Lorentz tensor structure of the considered amplitude.
We follow the method introduced in refs.~\cite{Peraro:2019cjj,Peraro:2020sfm} which has been used for the same tensors before in refs.~\cite{Bargiela:2021wuy,Caola:2021izf,Bargiela:2022lxz}.
We work in the 't Hooft-Veltman (tHV) scheme~\cite{THOOFT1972189} of dimensional regularization (dimReg) in $d=4-2\ep$ with $\ep \to 0$.
This means that we keep all loop momenta in $d$ dimensions while the external momenta $p_i^\mu$ are purely four dimensional.
As explained in refs.~\cite{Peraro:2019cjj,Peraro:2020sfm}, in this scheme, the amplitude can be decomposed into the same number of tensors as there are helicity configurations.
Up to overall parity, we have 8 such configurations, so the (bare) amplitude exhibits the following Lorentz structure
\begin{equation}\label{eq:hel_1}
  \mathcal{M}_{\rm bare} = (8 \alpha_b^2)\, A\,, \qquad\qquad A = \sum_{i=1}^8 \oF_i \, \oT_i \,,
\end{equation}
where we refer to
$\oF_i$
as form factors and to
$\oT_i$
as tensors. Note that above we have factored out the overall LO power of the (bare) electromagnetic coupling constant. 

Setting the reference vectors for the external polarization vectors $\epsilon_i^\mu(p_i,q_i)$ to $q_i^\mu = p_{i+1}^\mu$ (we identify $p_5\equiv p_1$), we obtain the following conditions
\begin{align}
\epsilon_i \cdot p_{1} = \epsilon_i \cdot p_{i+1} = 0 \,,\;\; \mbox{for} \;\; i=1,...,4 \,,
\label{eq:gauge}
\end{align}
which yield the tensor basis 
\begin{align}
\oT_1 &= p_1\!\cdot\!\ep_2 \; p_1\!\cdot\!\ep_3 \; p_2\!\cdot\!\ep_4 \; p_3\!\cdot\!\ep_1, \notag\\
\oT_2 &= \ep_3\!\cdot\!\ep_4 \; p_1\!\cdot\!\ep_2 \; p_3\!\cdot\!\ep_1 , \quad
\oT_3 = \ep_2\!\cdot\!\ep_4 \; p_1\!\cdot\!\ep_3 \; p_3\!\cdot\!\ep_1 , \notag\\
\oT_4 &= \ep_2\!\cdot\!\ep_3 \; p_2\!\cdot\!\ep_4 \; p_3\!\cdot\!\ep_1 , \quad
\oT_5 = \ep_1\!\cdot\!\ep_4 \; p_1\!\cdot\!\ep_2 \; p_1\!\cdot\!\ep_3 , \notag\\
\oT_6 &= \ep_1\!\cdot\!\ep_3 \; p_1\!\cdot\!\ep_2 \; p_2\!\cdot\!\ep_4 , \quad
\oT_7 = \ep_1\!\cdot\!\ep_2 \; p_1\!\cdot\!\ep_3 \; p_2\!\cdot\!\ep_4 , \notag\\
\oT_8 &= \ep_1\!\cdot\!\ep_2 \; \ep_3\!\cdot\!\ep_4 + \ep_1\!\cdot\!\ep_4 \; \ep_2\!\cdot\!\ep_3 + \ep_1\!\cdot\!\ep_3 \; \ep_2\!\cdot\!\ep_4 .
\label{eq:ten}
\end{align}
The form factors $\oF_i$ can be extracted from the amplitude $A$ with appropriately constructed projectors defined via $\sum_{pol} P_j \oT_i = \delta_{ji}$.
In order to construct helicity amplitudes $A_\lambdavec$ from the form factors $\oF_i$, we evaluate the tensors $\oT_i$ on fixed helicity states $\lambdavec = \{ \lambda_1, \lambda_2, \lambda_3, \lambda_4\}$ 
\begin{equation}\label{eq:hel_2}
  A_\lambdavec = \sum_{i=1}^8 \oF_{i} \, \left. \oT_{i}\right|_\lambdavec \,.
\end{equation}
In order to explicitly fix the helicity states of the external photons, we employ the spinor helicity formalism (see \textit{e.g.} \cite{Dixon:1996wi}). In our all-incoming convention for the external momenta, we define the left-handed (negative helicity) spinors $|i\rangle$ and $\langle i|$ and the right-handed (positive helicity) ones $|i]$ and $[i|$. In this notation the polarization vectors take the form
\begin{equation}
\epsilon^\mu_{j,-}(p_j) = \frac{\langle p_j | \gamma^\mu | q_j ] }{ \sqrt{2} [ p_j q_j ]}\,, \quad
\epsilon^\mu_{j,+}(p_j) = \frac{\langle q_j | \gamma^\mu | p_j ] }{ \sqrt{2} \langle q_j p_j \rangle } \,.
\end{equation}
As a result, the helicity amplitude $A_\lambdavec$ on each of the 8 helicity states $\lambdavec$ decomposes as
\begin{equation}\label{eq:hel_3}
  A_{\lambdavec} = \mathcal{S}_{\lambdavec} \, f_{\lambdavec} \,,
\end{equation}
into spinor-dependent phases
\begin{align}
 \mathcal{S}_{++++} &= \frac{[1 2][3 4]}{\langle1 2\rangle\langle3
    4\rangle} \,, & 
    \mathcal{S}_{-+++} &=
  \frac{\langle1 2\rangle\langle1 4\rangle[2 4]}{\langle3
    4\rangle\langle2 3\rangle\langle2 4\rangle} \,, &
  \mathcal{S}_{+-++} &= \frac{\langle2 1\rangle\langle2 4\rangle[1
      4]}{\langle3 4\rangle\langle1 3\rangle\langle1 4\rangle} \,,
  \notag\\
  \mathcal{S}_{++-+} &= \frac{\langle3 2\rangle\langle3 4\rangle[2
      4]}{\langle1 4\rangle\langle2 1\rangle\langle2 4\rangle} \,,&
  \mathcal{S}_{+++-} &= \frac{\langle4 2\rangle\langle4 3\rangle[2
      3]}{\langle1 3\rangle\langle2 1\rangle\langle2 3\rangle} \,,&
   \mathcal{S}_{--++} &= \frac{\langle1 2\rangle[3 4]}{[1
      2]\langle3 4\rangle} \,,
   \notag\\ 
  \mathcal{S}_{-+-+} &= \frac{\langle1 3\rangle[2 4]}{[1 3]\langle2
    4\rangle} \,,&
    \mathcal{S}_{+--+} &= \frac{\langle2 3\rangle[1 4]}{[2 3]\langle1 4\rangle} \,,&&
\label{eq:helamp}
\end{align}
and spinor-independent factors
\begin{align}
 f_{++++} &=  \frac{t^2}{4}\left(\frac{2\oF_{6}}{u}-\frac{2\oF_{3}}{s}-\oF_{1}\right)+\oF_{8}\left(\frac{s}{u}+\frac{u}{s}+4\right)+\frac{t}{2}(\oF_{2}-\oF_{4}+\oF_{5}-\oF_{7})\,, \notag\\ 
 f_{-+++} &=  \,\,\,\, \frac{t^2}{4}\left(\frac{2\oF_{3}}{s}+\oF_{1}\right)+t\left(\frac{\oF_{8}}{s}+\frac{1}{2}(\oF_{4}+\oF_{6}-\oF_{2})\right)\,, \notag\\ 
 f_{+-++} &=  -\frac{t^2}{4}\left(\frac{2\oF_{6}}{u}-\oF_{1}\right)+t\left(\frac{\oF_{8}}{u}-\frac{1}{2}(\oF_{2}+\oF_{3}+\oF_{5})\right)\,, \notag\\ 
 f_{++-+} &= \,\,\,\, \frac{t^2}{4}\left(\frac{2\oF_{3}}{s}+\oF_{1}\right)+t\left(\frac{\oF_{8}}{s}+\frac{1}{2}(\oF_{6}+\oF_{7}-\oF_{5})\right)\,, \notag\\ 
 f_{+++-} &=  -\frac{t^2}{4}\left(\frac{2\oF_{6}}{u}-\oF_{1}\right)+t\left(\frac{\oF_{8}}{u}+\frac{1}{2}(\oF_{4}+\oF_{7}-\oF_{3})\right)\,, \notag\\ 
 f_{--++} &=  -\frac{t^2}{4}\oF_{1}+\frac{1}{2}t(\oF_{2}+\oF_{3}-\oF_{6}-\oF_{7})+2\oF_{8}\,, \notag\\ 
 f_{-+-+} &=  t^2\left(\frac{\oF_{8}}{su}-\frac{\oF_{3}}{2s}+\frac{\oF_{6}}{2u}-\frac{\oF_{1}}{4}\right)\,, \notag\\ 
 f_{+--+} &=  -\frac{t^2}{4}\oF_{1}+\frac{1}{2}t(\oF_{3}-\oF_{4}+\oF_{5}-\oF_{6})+2\oF_{8}
\,.
\label{eq:alphabeta}
\end{align}
Note that the remaining 8 parity-related helicity amplitudes can be obtained from the above set through
\begin{equation}
A_{-\lambdavec} =
A_{\lambdavec}|_{\langle ij \rangle \leftrightarrow [ji]} \,.
\end{equation}
In addition, due to Bose symmetry among all the four external photons, only 3 helicity amplitudes remain independent.
Choosing one of each all-plus, single-minus, and double-minus independent configurations to be $A_{++++}$, $A_{-+++}$, and $A_{--++}$, the remaining 5 helicity amplitudes can be computed as
\begin{equation}
\begin{aligned}
A_{+-++} &= A_{-+++}|_{ ( p_1 \leftrightarrow p_2, \lambda_1 \leftrightarrow \lambda_2 ) } \,, \\ 
A_{++-+} &= A_{-+++}|_{ ( p_1 \leftrightarrow p_3, \lambda_1 \leftrightarrow \lambda_3 ) } \,, \\ 
A_{+++-} &= A_{-+++}|_{ ( p_1 \leftrightarrow p_4, \lambda_1 \leftrightarrow \lambda_4 ) } \,, \\ 
A_{-+-+} &= A_{--++}|_{ ( p_2 \leftrightarrow p_3, \lambda_2 \leftrightarrow \lambda_3 ) } \,, \\ 
A_{+--+} &= A_{--++}|_{ ( p_1 \leftrightarrow p_3, \lambda_1 \leftrightarrow \lambda_3 ) } \,.
\end{aligned}
\label{eq:Bose}
\end{equation}
Bose symmetry also imposes a strong relation on single-minus little group scalars
\begin{equation}
f_{-+++} = f_{+-++} = f_{++-+} = f_{+++-} \,.
\end{equation}
Indeed, as $A_{-+++}$ is symmetric under three separate exchanges of particles, \textit{i.e.} 2 and 3, 3 and 4, as well as 2 and 4, and since its phase factor $\mathcal{S}_{-+++}$ preserves this symmetries, then $f_{-+++}$ must be also symmetric under each of these exchanges.
It is easy to note that after applying these three separate exchanges of momenta in the Mandelsam variable dependence of $f_{-+++}$, one obtains $f_{+++-}$, $f_{+-++}$, and $f_{++-+}$, respectively, according to eq.~\eqref{eq:Bose}.

Before presenting our results, we elaborate here on the details of our calculation of the bare amplitudes to three-loop order.
We start by generating all the relevant Feynman diagrams with \texttt{Qgraf}~\cite{Nogueira:1991ex}.
There are 6, 72, and 1296 diagrams at one-, two-, and three-loop order, respectively.
Then, we perform all the colour and Lorentz tensor algebra in \texttt{FORM}~\cite{Vermaseren:2000nd}.
Next, we map all of the resulting scalar Feynman integrals onto the integral topologies defined in ref.~\cite{Bargiela:2021wuy}.
Note that the nonplanar topology starts contributing at three-loop order.
We reduce the number of the integrals to be computed by about one order of magnitude by removing scaleless integrals, which are all zero in dimReg.
We further reduce the number of these integrals by around a factor of two by mapping the integrals into common topology sectors.

After evaluating the Dirac traces due to the internal fermion lines, the amplitude can be expressed in terms of
scalar Feynman integrals. The most demanding step is to express these
integrals in terms of a basis of master integrals (MIs) using
integration-by-parts (IBP) identities~\cite{Chetyrkin:1981qh}. For this
purpose we employ the Laporta algorithm~\cite{Laporta:2000dsw}, as
implemented in \texttt{Reduze 2}~\cite{vonManteuffel:2012np,Studerus:2009ye}
and the in-house reduction framework \texttt{Finred}~\cite{vonManteuffel:2016xki}.
These tools incorporate modern developments in IBP reduction techniques,
including approaches based on syzygy relations
\cite{Gluza:2010ws,Ita:2015tya,Larsen:2015ped,Bohm:2017qme,Schabinger:2011dz,Agarwal:2020dye}
and finite-field reconstruction methods
\cite{vonManteuffel:2014ixa,Peraro:2016wsq,Peraro:2019svx,vonManteuffel:2016xki}.

At the three-loop order we reduced 1046489 integrals to a basis of 486 master integrals.
For a large fraction of the integrals we were able to reuse reduction
tables obtained in earlier three-loop four-point computations.
However, an additional set of 11,297 integrals was not covered by the
existing tables. These missing reductions were generated specifically
for this work using \texttt{Reduze 2} and \texttt{Kira 3}~\cite{Lange:2025fba}, after which
they were incorporated into the existing reduction database.

Direct substitution of the full set of IBP relations is computationally
expensive, as intermediate expressions can grow to sizes of order of 500 GB. To control this growth, the reductions are performed in a
staged manner. In each step a selected subset of IBP relations is
inserted, followed by simplification of the resulting expressions before
the next subset is applied. This sequential strategy keeps intermediate
files manageable and avoids severe combinatorial growth. The subsets of
relations are chosen such that the number of substitutions and the
complexity of the resulting denominators remain balanced. Additional
simplification procedures are applied immediately after each stage in
order to prevent uncontrolled expansion of the expressions.

A major source of expression growth during IBP substitution is the
appearance of complicated denominator structures. Performing a global
partial fraction decomposition from the outset is often inefficient and
computationally expensive. To address this issue we implemented a
dedicated Mathematica routine based on
\texttt{MultivariateApart}~\cite{Heller:2021qkz}. In this approach,
each rational coefficient is split into terms, each having a unique denominator. The possible denominators include the trivial one (1), as well as products of various factors. Each term in the rational coefficients is then factorised and
partial fractioned separately and, after all terms have been simplified, a final global partial fraction decomposition\footnote{Here rather than a simple sum, we require a final step involving multivariate partial fractioning with respect to a common Gr\"obner basis, since each term has been previously partial fractioned on its own Gr\"obner basis.} is performed to assemble the complete result.

Handling denominators in this grouped manner avoids large global
expansions and makes common factors more transparent, which
significantly reduces the total number of generated terms. The routine
is applied after each IBP stage so that subsequent reductions start
from already simplified expressions. Without such targeted partial
fraction decomposition, further IBP insertions might rapidly lead to
expressions that are computationally intractable. To ensure the correctness of the simplifications, numerical cross-checks
were performed at several intermediate stages of the calculation. The
simplified expressions were evaluated for non-trivial numerical values
of the colour factors and the dimensional parameter $d$, and compared
with the corresponding unsimplified expressions.

The three-loop MIs have been computed before in refs.~\cite{Henn:2020lye,Bargiela:2021wuy}.
All the required MIs up to three loops are expressed in terms of Harmonic Polylogarithms (HPLs)
\begin{equation}\label{eq:HPL}
  G(\underbrace{0,\dots,0}_{n~\text{times}};x) \equiv \frac{\ln^n x }{n!},~~~~~~~
G(a_n,...,a_1;x) = \int_0^x \frac{dz}{z-a_n} G(a_{n-1},...,a_1;z),
\end{equation}
with $G(x)=1$ and letters $a_i\in\{0,1\}$ at each order in the dimensional regulator $\ep$.
In order to extract the finite part of the three-loop amplitude, it is enough to truncate the expansion in powers of $\ep$ at transcendental weight 6.
We used \texttt{PolyLogTools}~\cite{Duhr:2019tlz} and our in-house routines to manipulate HPLs.

The transcendental functions defined above are real valued in the ``physical region'' defined by $1>x>0$. However, when performing the crossings of external momenta required by \cref{eq:Bose}, one might exit the physical region and cross one or more branch cuts of the HPLs, which therefore need to be analytically continued. Thanks to the shuffle algebra  relations satisfied by all Multiple Polylogarithms (MPLs)\footnote{HPLs are a subset of MPLs.} (see \textit{e.g.} refs.~\cite{Duhr:2014woa,Weinzierl:2022eaz}) 
\def\dsqcup{\sqcup\mathchoice{\mkern-7mu}{\mkern-7mu}{\mkern-3.2mu}{\mkern-3.8mu}\sqcup}
\begin{equation}
    G(\vec{a};x)\,G(\vec{b};x) = \sum_{\vec{s} \, \in \, \vec{a} \,\dsqcup\, \vec{b} } G(\vec{s};x)\,,
\end{equation}
the only information we need in order to perform the analytic continuation of HPLs is that of the weight-1 ones, \textit{i.e.} the logarithms: 
\begin{equation}
    \begin{aligned}
        p_1 \leftrightarrow p_2\; (p_3 \leftrightarrow p_4) &: \quad
        \log(x) \to \log(1-x) \,, \quad 
        &&\log(1-x) \to \log(x)\, , \\
        p_1 \leftrightarrow p_3\; (p_2 \leftrightarrow p_4) &: \quad    
        \log(x) \to \log\left(\frac{x}{1-x}\right) -i \pi\,, \quad 
        &&\log(1-x) \to -\log(1-x) -2 i \pi\, , \\
        p_2 \leftrightarrow p_3\; (p_1 \leftrightarrow p_4) &: \quad    
        \log(x) \to -\log(x) -2 i \pi\,, \quad 
        &&\log(1-x) \to \log\left(\frac{1-x}{x}\right) -i \pi\, .
    \end{aligned}
\end{equation}

The finite part of the amplitude is constructed from the bare amplitude by analysing its ultraviolet (UV) and infrared (IR) properties.
The perturbative series for the bare spinor-stripped helicity amplitudes up to the three-loop order in QCD reads
\begin{equation}
f_{\vec\lambda} =  f^{(1)}_{\vec\lambda} + \left(\frac{\alpha_{s,b}}{\pi}\right) \, f^{(2)}_{\vec\lambda} + \left(\frac{\alpha_{s,b}}{\pi}\right)^2 \, f^{(3)}_{\vec\lambda} + \mathcal{O}(\alpha_{s,b}^3)  \,,
\end{equation}
where the superscript in $f^{(L)}_{\vec\lambda}$ denotes the number of loops in the bare amplitudes.

Because $f_{\vec\lambda}^{(1)}$ is LO in QED and $f_{\vec\lambda}^{(2)}$ is LO both in QED and QCD, they are both free of UV singularities. Furthermore, the QCD IR anomalous dimension for $n$-point photon scattering is vanishing,\footnote{The IR anomalous dimension is vanishing for LbL, provided that renormalisation is performed in the scheme discussed in \cref{sec:ren}.} which ensures that LbL scattering is free of IR divergences. This in particular implies that bare amplitudes $f_{\vec\lambda}^{(1)}$ and $f_{\vec\lambda}^{(2)}$ do not have any $\frac{1}{\ep^n}$ poles.
In contrast to this, the three-loop bare amplitude has a single UV pole, \textit{i.e.} $f_{\vec\lambda}^{(3)} \sim \mathcal{O}\left(\frac{1}{\ep}\right)$. 

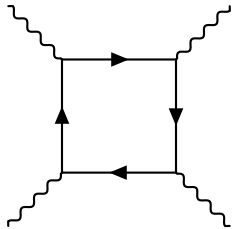
\begin{figure}[t]
    \centering
    \begin{equation*}
\begin{gathered}
    \begin{tikzpicture}[line width=0.8,scale=1,line cap=round]
          \begin{feynman}
            \vertex (a);
            \vertex [below=of a] (b);
            \vertex [left=of b] (c);
            \vertex [above=of c] (d);
           \vertex [above right=1cm of a] (a1);
            \vertex [below right=1cm of b] (b1);
            \vertex [below left=1cm of c] (c1);
            \vertex [above left=1cm of d] (d1);

            \diagram* {
                (a) -- [fermion] (b),
                (b) -- [fermion] (c),
                (c) -- [fermion] (d),
                (d) -- [fermion] (a),
                (a) -- [photon] (a1),
                (b) -- [photon] (b1),
                (c) -- [photon] (c1),
                (d) -- [photon] (d1),
                            };
            \end{feynman}
    \end{tikzpicture}
\end{gathered}
\end{equation*}
    \caption{The only diagram type contributing to the one-loop amplitude.}
    \label{fig:one-loop}
\end{figure}

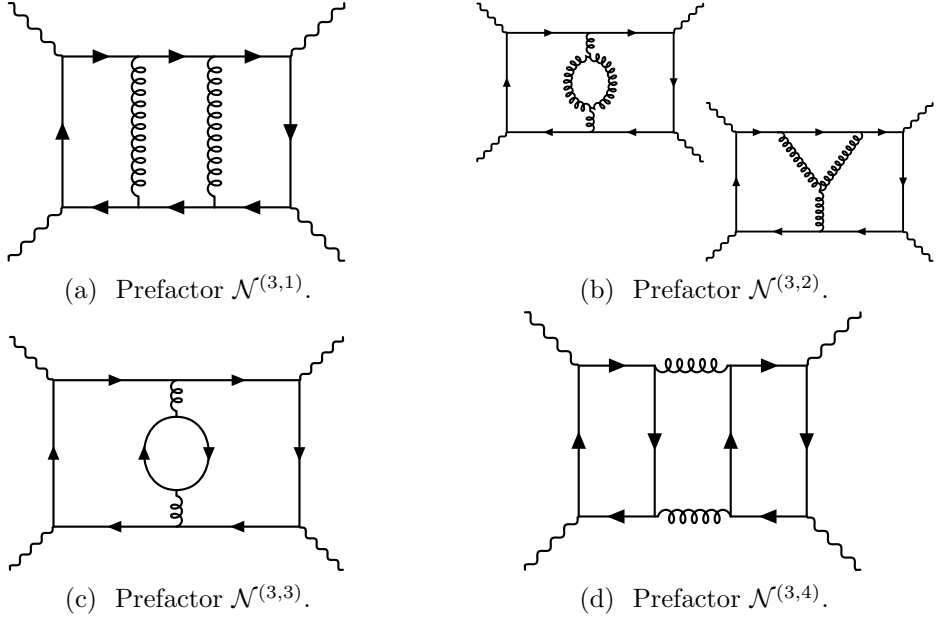
\begin{figure}[t]
    \centering
    \begin{subfigure}[b]{0.3\textwidth}
        \centering
        \begin{equation*}
\hspace{-0.3cm}
\vspace{-0.6cm}
\begin{gathered}
    \begin{tikzpicture}[line width=0.8,scale=1,line cap=round]
          \begin{feynman}
            \vertex (a);
            \vertex [below=2cm of a] (b);
            \vertex [left=1cm of b] (y);
            \vertex [left=1cm of y] (z);
            \vertex [left=1cm of z] (c);
            \vertex [above=2cm of c] (d);
            \vertex [right=1cm of d] (x);
            \vertex [right=1cm of x] (w);
           \vertex [above right=1cm of a] (a1);
            \vertex [below right=1cm of b] (b1);
            \vertex [below left=1cm of c] (c1);
            \vertex [above left=1cm of d] (d1);

            \diagram* {
                (a) -- [fermion] (b),
                (b) -- [fermion] (y),
                (y) -- [fermion] (z),
                (z) -- [fermion] (c),
                (c) -- [fermion] (d),
                (d) -- [fermion] (x),
                (x) -- [fermion] (w),
                (w) -- [fermion] (a),
                (a) -- [photon] (a1),
                (b) -- [photon] (b1),
                (c) -- [photon] (c1),
                (d) -- [photon] (d1),
                (x) -- [gluon] (z),
                (w) -- [gluon] (y)
                            };
            \end{feynman}
    \end{tikzpicture}
\end{gathered}
\end{equation*}
        \caption{\label{fig:3loop1} Prefactor $\mathcal{N}^{(3,1)}$.}
    \end{subfigure}
    \hspace{2cm}
    \begin{subfigure}[b]{0.3\textwidth}
        \centering
        \begin{equation*}
\hspace{-0.8cm}
\vspace{-0.6cm}
\begin{gathered}
\resizebox{16em}{!}{
\begin{tikzpicture}[line width=1.3,scale=1,line cap=round]
      \begin{feynman}
        \vertex (a);
        \vertex [below=2.4cm of a] (b);
        \vertex [left=2cm of b] (y);
        \vertex [left=2cm of y] (c);
        \vertex [above=2.4cm of c] (d);
        \vertex [right=2cm of d] (x);
        \vertex [above right=1cm of a] (a1);
        \vertex [below right=1cm of b] (b1);
        \vertex [below left=1cm of c] (c1);
        \vertex [above left=1cm of d] (d1);
        \vertex [below=0.6cm of x] (x1);
        \vertex [above=0.6cm of y] (y1);

        \vertex [right=5.5cm of b] (aX);
        \vertex [below=2.4cm of aX] (bX);
        \vertex [left=2cm of bX] (yX);
        \vertex [left=2cm of yX] (cX);
        \vertex [above=2.4cm of cX] (dX);
        \vertex [right=1cm of dX] (x1X);
        \vertex [right=2cm of x1X] (x2X);
        \vertex [above right=1cm of aX] (a1X);
        \vertex [below right=1cm of bX] (b1X);
        \vertex [below left=1cm of cX] (c1X);
        \vertex [above left=1cm of dX] (d1X);
        \vertex [above=1cm of yX] (y1X);
        
        \diagram* {
            (a) -- [fermion] (b),
            (b) -- [fermion] (y),
            (y) -- [fermion] (c),
            (c) -- [fermion] (d),
            (d) -- [fermion] (x),
            (x) -- [fermion] (a),
            (a) -- [photon] (a1),
            (b) -- [photon] (b1),
            (c) -- [photon] (c1),
            (d) -- [photon] (d1),
            (x) -- [gluon] (x1),
            (y) -- [gluon] (y1),
            (x1) -- [gluon, half left] (y1),
            (y1) -- [gluon, half left] (x1),

            (aX) -- [fermion] (bX),
            (bX) -- [fermion] (yX),
            (yX) -- [fermion] (cX),
            (cX) -- [fermion] (dX),
            (dX) -- [fermion] (x1X),
            (x1X) -- [fermion] (x2X),
            (x2X) -- [fermion] (aX),
            (aX) -- [photon] (a1X),
            (bX) -- [photon] (b1X),
            (cX) -- [photon] (c1X),
            (dX) -- [photon] (d1X),
            (yX) -- [gluon] (y1X),
            (y1X) -- [gluon] (x1X),
            (y1X) -- [gluon] (x2X)

                        };
        \end{feynman}
    \end{tikzpicture}
}
\end{gathered}
\end{equation*}
        \caption{\label{fig:3loop4a} Prefactor $\mathcal{N}^{(3,2)}$.}
    \end{subfigure}
    \begin{subfigure}[b]{0.3\textwidth}
        \centering
        \begin{equation*}
\hspace{-0.3cm}
\vspace{-0.6cm}
    \begin{gathered}
    \resizebox{12em}{!}{
    \begin{tikzpicture}[line width=1,scale=1,line cap=round]
          \begin{feynman}
            \vertex (a);
            \vertex [below=2.4cm of a] (b);
            \vertex [left=2cm of b] (y);
            \vertex [left=2cm of y] (c);
            \vertex [above=2.4cm of c] (d);
            \vertex [right=2cm of d] (x);
            \vertex [above right=1cm of a] (a1);
            \vertex [below right=1cm of b] (b1);
            \vertex [below left=1cm of c] (c1);
            \vertex [above left=1cm of d] (d1);
            \vertex [below=0.6cm of x] (x1);
            \vertex [above=0.6cm of y] (y1);
            
            \diagram* {
                (a) -- [fermion] (b),
                (b) -- [fermion] (y),
                (y) -- [fermion] (c),
                (c) -- [fermion] (d),
                (d) -- [fermion] (x),
                (x) -- [fermion] (a),
                (a) -- [photon] (a1),
                (b) -- [photon] (b1),
                (c) -- [photon] (c1),
                (d) -- [photon] (d1),
                (x) -- [gluon] (x1),
                (y) -- [gluon] (y1),
                (x1) -- [fermion, half left] (y1),
                (y1) -- [fermion, half left] (x1)
                            };
            \end{feynman}
    \end{tikzpicture}
    }
\end{gathered}
\end{equation*}
        \caption{\label{fig:3loop3} Prefactor $\mathcal{N}^{(3,3)}$.}
    \end{subfigure}
    \hspace{2cm}
    \begin{subfigure}[b]{0.3\textwidth}
        \centering
        \begin{equation*}
\hspace{-0.3cm}
\vspace{-0.6cm}
    \begin{gathered}
    \begin{tikzpicture}[line width=0.8,scale=1,line cap=round]
          \begin{feynman}
            \vertex (a);
            \vertex [below=2cm of a] (b);
            \vertex [left=1cm of b] (y);
            \vertex [left=1cm of y] (z);
            \vertex [left=1cm of z] (c);
            \vertex [above=2cm of c] (d);
            \vertex [right=1cm of d] (x);
            \vertex [right=1cm of x] (w);
           \vertex [above right=1cm of a] (a1);
            \vertex [below right=1cm of b] (b1);
            \vertex [below left=1cm of c] (c1);
            \vertex [above left=1cm of d] (d1);

            \diagram* {
                (a) -- [fermion] (b),
                (b) -- [fermion] (y),
                (y) -- [gluon] (z),
                (z) -- [fermion] (c),
                (c) -- [fermion] (d),
                (d) -- [fermion] (x),
                (x) -- [gluon] (w),
                (w) -- [fermion] (a),
                (a) -- [photon] (a1),
                (b) -- [photon] (b1),
                (c) -- [photon] (c1),
                (d) -- [photon] (d1),
                (x) -- [fermion] (z),
                (y) -- [fermion] (w)
                            };
            \end{feynman}
    \end{tikzpicture}
\end{gathered}
\end{equation*}
        \caption{\label{fig:3loop2} Prefactor $\mathcal{N}^{(3,4)}$.}
    \end{subfigure} 
    \caption{Representative diagrams for the gauge invariant parts of the three-loop amplitude. Other members in each set are obtained by permutations of the external momenta and of the attachments of the internal gluons to the fermion lines. Diagrams contributing to $\mathcal{N}^{(3,2)}$ are purely non-abelian and therefore give no contribution to the QED corrections to the amplitude.}
    \label{fig:three-loop}
\end{figure}

In addition to the decomposition in helicity amplitudes, one can further split these into smaller gauge invariant building blocks ${{f}}^{(i,j)}_{\vec\lambda}$ identified by their the colour and flavour structures: 
\begin{equation} \label{eq:colour_and_flavour_decomp}
    {f}^{(i)}_{\vec\lambda} = \sum_{j} \mathcal{N}^{(i,j)} \, {{f}}^{(i,j)}_{\vec\lambda} \,.
\end{equation}
At one loop we find a single term given by
\begin{equation}
    \mathcal{N}^{(1,1)} = N_c \,\tilde{n}_q^{(4)} \,.
\end{equation}
This prefactor is associated to the diagram in \cref{fig:one-loop}, which apart from permutations of the external momenta is the only diagram appearing at this perturbative order.
At two loops we again find a single colour and flavour prefactor, shared by all diagrams of \cref{fig:two-loop}. Explicitly it reads
\begin{equation}
    \mathcal{N}^{(2,1)} = N_c\, C_F \, \tilde{n}_q^{(4)}  \,.
\end{equation}
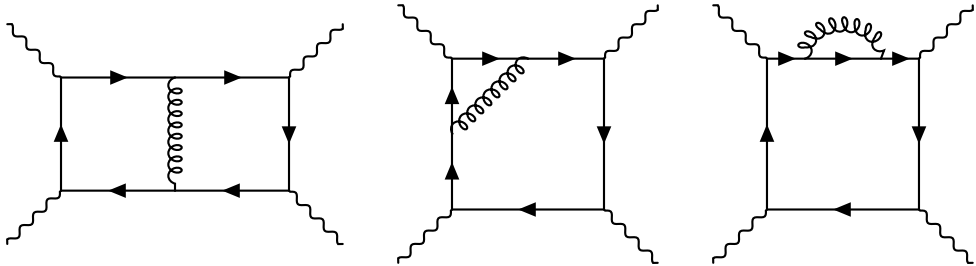
\begin{figure}[t]
    \centering
    \begin{equation*}
\begin{gathered}
    \begin{tikzpicture}[line width=0.8,scale=1,line cap=round]
          \begin{feynman}
            \vertex (a);
            \vertex [below=of a] (b);
            \vertex [left=of b] (y);
            \vertex [left=of y] (c);
            \vertex [above=of c] (d);
            \vertex [right=of d] (x);
            \vertex [above right=1cm of a] (a1);
            \vertex [below right=1cm of b] (b1);
            \vertex [below left=1cm of c] (c1);
            \vertex [above left=1cm of d] (d1);

            \diagram* {
                (a) -- [fermion] (b),
                (b) -- [fermion] (y),
                (y) -- [fermion] (c),
                (c) -- [fermion] (d),
                (d) -- [fermion] (x),
                (x) -- [fermion] (a),
                (a) -- [photon] (a1),
                (b) -- [photon] (b1),
                (c) -- [photon] (c1),
                (d) -- [photon] (d1),
                (x) -- [gluon] (y),
                            };
            \end{feynman}
    \end{tikzpicture}
\end{gathered}
\hspace{0.7cm}
\begin{gathered}
    \begin{tikzpicture}[line width=0.8,scale=1,line cap=round]
          \begin{feynman}
            \vertex (a);
            \vertex [below=2cm of a] (b);
            \vertex [left=2cm of b] (c);
            \vertex [above=1cm of c] (y);
            \vertex [above=1cm of y] (d);
            \vertex [right=1cm of d] (x);
            \vertex [above right=1cm of a] (a1);
            \vertex [below right=1cm of b] (b1);
            \vertex [below left=1cm of c] (c1);
            \vertex [above left=1cm of d] (d1);

            \diagram* {
                (a) -- [fermion] (b),
                (b) -- [fermion] (c),
                (c) -- [fermion] (y),
                (y) -- [fermion] (d),
                (d) -- [fermion] (x),
                (x) -- [fermion] (a),
                (a) -- [photon] (a1),
                (b) -- [photon] (b1),
                (c) -- [photon] (c1),
                (d) -- [photon] (d1),
                (x) -- [gluon] (y),
                            };
            \end{feynman}
    \end{tikzpicture}
\end{gathered}
\hspace{0.7cm}
\begin{gathered}
    \begin{tikzpicture}[line width=0.8,scale=1,line cap=round]
          \begin{feynman}
            \vertex (a);
            \vertex [below=2cm of a] (b);
            \vertex [left=2cm of b] (c);
            \vertex [above=2cm of c] (d);
            \vertex [right=0.5cm of d] (x);
            \vertex [right=1cm of x] (y);
            \vertex [above right=1cm of a] (a1);
            \vertex [below right=1cm of b] (b1);
            \vertex [below left=1cm of c] (c1);
            \vertex [above left=1cm of d] (d1);

            \diagram* {
                (a) -- [fermion] (b),
                (b) -- [fermion] (c),
                (c) -- [fermion] (d),
                (d) -- [fermion] (x),
                (x) -- [fermion] (y),
                (y) -- [fermion] (a),
                (a) -- [photon] (a1),
                (b) -- [photon] (b1),
                (c) -- [photon] (c1),
                (d) -- [photon] (d1),
                (x) -- [gluon, half left] (y),
                            };
            \end{feynman}
    \end{tikzpicture}
\end{gathered}
\end{equation*}
    \caption{Representative diagrams for the two-loop amplitude. All other diagrams can be obtained via permutations of the external legs. All diagrams in the two-loop amplitude correspond to the same colour and flavour factors and are proportional to $\alpha^2 \alpha_s N_c C_F \tilde{n}_q^{(4)}$.}
    \label{fig:two-loop}
\end{figure} 
At the three-loop level, the amplitude involves four gauge-invariant subsets defined by
\begin{equation}\label{eq:three-loop-colour-def-QCD}
\begin{aligned}
    \mathcal{N}^{(3,1)} &= N_c\, C_F^2 \, \tilde{n}_q^{(4)} \,,
    &&
    \mathcal{N}^{(3,2)} = N_c\, C_F\, C_A\, \tilde{n}_q^{(4)}  \,,
    \\
    \mathcal{N}^{(3,3)} &= N_c\, C_F\, T_F\, \tilde{n}_q^{(4)}\, n_q  \,,
    &&
    \mathcal{N}^{(3,4)} = N_c \, C_F\, T_F\, \left(\tilde{n}_q^{(2)}\right)^2 \,,
\end{aligned}
\end{equation}
and which are associated to the diagrams in \cref{fig:three-loop}.

Having defined the gauge invariant building blocks $f^{(i,j)}_{\vec\lambda}$ in such a way that their colour factors correspond to specific combinations of $SU(N_c)$ Casimir invariants, we now describe how to abelianise the QCD results in order to obtain QED and QED-QCD mixed bare amplitudes up to three loops.

\section{QED corrections and abelianisation}
\label{sec:qed}
Once the pure QCD corrections corresponding to the gauge-invariant subamplitudes ${f}^{(i,j)}_{\vec\lambda}$ have been computed, one can extract the QED as well as the mixed QED-QCD ones in a straightforward way. For further discussions on the abelianisation of QCD amplitudes, we guide the reader to refs.~\cite{deFlorian:2018wcj,AH:2019pyp}. We write the full\footnote{Here by full we mean the ``QCD + QED + mixed'' amplitudes.} bare spinor-stripped helicity amplitudes up to three loops as
\begin{equation}\label{eq:full_bare_amplitude}
\begin{split}
    {f}_{\text{QCD+QED},\vec\lambda} &= \widetilde{\mathcal{N}}^{(1,1)} \, {{f}}^{(1,1)}_{\vec\lambda} +
    \widetilde{\mathcal{N}}^{(2,1)} \, {{f}}^{(2,1)}_{\vec\lambda} + 
    \sum_{j=1}^4\widetilde{\mathcal{N}}^{(3,j)} \, {{f}}^{(3,j)}_{\vec\lambda} \,,
\end{split}
\end{equation}
where the new prefactors $\widetilde{\mathcal{N}}^{(i,j)}$ now include powers of $\alpha_{s,b}$ as well as of the bare electromagnetic coupling $\alpha_b = e_b^2/4\pi$ and are defined as
\begin{equation}\label{eq:three-loop-colour-def-MIXED}
\begin{aligned}
    \widetilde{\mathcal{N}}^{(1,1)} &= n_l + N_c\, \tilde{n}_q^{(4)} \,,
    \\
    \widetilde{\mathcal{N}}^{(2,1)} &=  \left(\frac{\alpha_b}{\pi}\right) \left(n_l + N_c\, \tilde{n}_q^{(6)} \right) + \left( \frac{\alpha_{s,b}}{\pi} \right) \, N_c\, C_F \, \tilde{n}_q^{(4)} \,,
    \\
    \widetilde{\mathcal{N}}^{(3,1)} &= \left(\frac{\alpha_b}{\pi}\right)^2 \left(n_l + N_c \, \tilde{n}_q^{(8)} \right) + \left(\frac{\alpha_b}{\pi}\right) \!\left(\frac{\alpha_{s,b}}{\pi}\right) 2\, N_c\, C_F\, \tilde{n}_q^{(6)} + \left(\frac{\alpha_{s,b}}{\pi}\right)^2  N_c\, C_F^2 \, \tilde{n}_q^{(4)}  \,,
    \\
    \widetilde{\mathcal{N}}^{(3,2)} &=  \left(\frac{\alpha_{s,b}}{\pi}\right)^2 \, N_c\, C_F\, C_A\, \tilde{n}_q^{(4)}  \,,
    \\
    \widetilde{\mathcal{N}}^{(3,3)} &=  \left(\frac{\alpha_b}{\pi}\right)^2 \left(n_l + N_c\, \tilde{n}_q^{(6)} \right) \left(n_l + N_c\, \tilde{n}_q^{(2)} \right)  + \left(\frac{\alpha_{s,b}}{\pi}\right)^2 \, N_c\, C_F\, T_F\, \tilde{n}_q^{(4)}\, n_q \,,
    \\
    \widetilde{\mathcal{N}}^{(3,4)} &= \left(\frac{\alpha_b}{\pi}\right)^2 \left(n_l + N_c\, \tilde{n}_q^{(4)} \right)^2 + \left(\frac{\alpha_{s,b}}{\pi}\right)^2 \, N_c \, C_F\, T_F\, \left(\tilde{n}_q^{(2)}\right)^2 \,.
 \end{aligned}
 \end{equation}
Above we have used $n_l$ to refer to the number of active charged lepton flavours.
The spinor-stripped helicity amplitudes ${f}_{\text{QCD+QED},\vec\lambda}^{(i)}$ contain all QCD and/or (massless) QED diagrams up to three loops. The corresponding colour and flavour coefficients $\widetilde{\mathcal{N}}^{(i,j)}$ can be obtained from replacing one, two or none of the gluons in \cref{fig:one-loop,fig:two-loop,fig:three-loop} with photons, allowing both massless quarks and leptons to run in the loops and computing the corresponding kinematics-independent factor. 
For instance, for the diagram in \cref{fig:3loop1} we find 
\begin{equation}
\begin{aligned}
    &\resizebox{7em}{!}{
    \begin{tikzpicture}[baseline=-1cm, line width=1,scale=1,line cap=round]
        \begin{feynman}
            \vertex (a);
            \vertex [below=2cm of a] (b);
            \vertex [left=1cm of b] (y);
            \vertex [left=1cm of y] (z);
            \vertex [left=1cm of z] (c);
            \vertex [above=2cm of c] (d);
            \vertex [right=1cm of d] (x);
            \vertex [right=1cm of x] (w);
            \vertex [above right=1cm of a] (a1);
            \vertex [below right=1cm of b] (b1);
            \vertex [below left=1cm of c] (c1);
            \vertex [above left=1cm of d] (d1);

            \diagram* {
                (a) -- [fermion] (b),
                (b) -- [fermion] (y),
                (y) -- [fermion] (z),
                (z) -- [fermion] (c),
                (c) -- [fermion] (d),
                (d) -- [fermion] (x),
                (x) -- [fermion] (w),
                (w) -- [fermion] (a),
                (a) -- [photon] (a1),
                (b) -- [photon] (b1),
                (c) -- [photon] (c1),
                (d) -- [photon] (d1),
                (x) -- [gluon] (z),
                (w) -- [gluon] (y)
                            };
            \end{feynman}
    \end{tikzpicture}
    } \propto \left( \frac{\alpha_{s,b}}{\pi}\right)^2  \,N_c  \,C_F^2  \,\tilde{n}_q^{(4)} \,, \quad\quad
    \resizebox{7em}{!}{
    \begin{tikzpicture}[baseline=-1cm, line width=1,scale=1,line cap=round]
        \begin{feynman}
            \vertex (a);
            \vertex [below=2cm of a] (b);
            \vertex [left=1cm of b] (y);
            \vertex [left=1cm of y] (z);
            \vertex [left=1cm of z] (c);
            \vertex [above=2cm of c] (d);
            \vertex [right=1cm of d] (x);
            \vertex [right=1cm of x] (w);
            \vertex [above right=1cm of a] (a1);
            \vertex [below right=1cm of b] (b1);
            \vertex [below left=1cm of c] (c1);
            \vertex [above left=1cm of d] (d1);

            \diagram* {
                (a) -- [fermion] (b),
                (b) -- [fermion] (y),
                (y) -- [fermion] (z),
                (z) -- [fermion] (c),
                (c) -- [fermion] (d),
                (d) -- [fermion] (x),
                (x) -- [fermion] (w),
                (w) -- [fermion] (a),
                (a) -- [photon] (a1),
                (b) -- [photon] (b1),
                (c) -- [photon] (c1),
                (d) -- [photon] (d1),
                (x) -- [photon] (z),
                (w) -- [photon] (y)
                            };
            \end{feynman}
    \end{tikzpicture}
    } \propto \left( \frac{\alpha_b}{\pi}\right)^2 \, \left( n_l + N_c   \,\tilde{n}_q^{(8)} \right)\,, \\
    &
    \quad\quad\quad\quad\quad
    \resizebox{7em}{!}{
    \begin{tikzpicture}[baseline=-1cm, line width=1,scale=1,line cap=round]
        \begin{feynman}
            \vertex (a);
            \vertex [below=2cm of a] (b);
            \vertex [left=1cm of b] (y);
            \vertex [left=1cm of y] (z);
            \vertex [left=1cm of z] (c);
            \vertex [above=2cm of c] (d);
            \vertex [right=1cm of d] (x);
            \vertex [right=1cm of x] (w);
            \vertex [above right=1cm of a] (a1);
            \vertex [below right=1cm of b] (b1);
            \vertex [below left=1cm of c] (c1);
            \vertex [above left=1cm of d] (d1);

            \diagram* {
                (a) -- [fermion] (b),
                (b) -- [fermion] (y),
                (y) -- [fermion] (z),
                (z) -- [fermion] (c),
                (c) -- [fermion] (d),
                (d) -- [fermion] (x),
                (x) -- [fermion] (w),
                (w) -- [fermion] (a),
                (a) -- [photon] (a1),
                (b) -- [photon] (b1),
                (c) -- [photon] (c1),
                (d) -- [photon] (d1),
                (x) -- [photon] (z),
                (w) -- [gluon] (y)
                            };
            \end{feynman}
    \end{tikzpicture}
    } 
    +
    \resizebox{7em}{!}{
    \begin{tikzpicture}[baseline=-1cm, line width=1,scale=1,line cap=round]
        \begin{feynman}
            \vertex (a);
            \vertex [below=2cm of a] (b);
            \vertex [left=1cm of b] (y);
            \vertex [left=1cm of y] (z);
            \vertex [left=1cm of z] (c);
            \vertex [above=2cm of c] (d);
            \vertex [right=1cm of d] (x);
            \vertex [right=1cm of x] (w);
            \vertex [above right=1cm of a] (a1);
            \vertex [below right=1cm of b] (b1);
            \vertex [below left=1cm of c] (c1);
            \vertex [above left=1cm of d] (d1);

            \diagram* {
                (a) -- [fermion] (b),
                (b) -- [fermion] (y),
                (y) -- [fermion] (z),
                (z) -- [fermion] (c),
                (c) -- [fermion] (d),
                (d) -- [fermion] (x),
                (x) -- [fermion] (w),
                (w) -- [fermion] (a),
                (a) -- [photon] (a1),
                (b) -- [photon] (b1),
                (c) -- [photon] (c1),
                (d) -- [photon] (d1),
                (x) -- [gluon] (z),
                (w) -- [photon] (y)
                            };
            \end{feynman}
    \end{tikzpicture}
    }
    \propto \left( \frac{\alpha_b}{\pi}\right) \left( \frac{\alpha_{s,b}}{\pi}\right)  \, 2 \,N_c  \,C_F  \,\tilde{n}_q^{(6)} \,, 
\end{aligned}
\end{equation}
where we have stripped out an overall $\alpha_b^2$ factor according to \cref{eq:hel_1}.
Summing all three contributions one obtains the value of $\widetilde{\mathcal{N}}^{(3,1)}$ given above. One proceeds in a similar fashion for the other contributions. Note that since the class of diagrams depicted in \cref{fig:3loop4a} is purely non-abelian, it does not generate any QED corrections. 

The bare helicity amplitudes computed in \cref{eq:full_bare_amplitude} contain simple $1/\epsilon$ poles of pure UV origin. Indeed, the renormalised IR anomalous dimension for LbL scattering defined this way vanishes exactly and (a suitably performed) renormalisation of the strong and electric coupling suffices to make the amplitudes fully finite. We give details about this in the next section.


\section{Renormalisation and finite remainders}
\label{sec:ren}
For the renormalisation of the amplitudes described above we follow ref.~\cite{Bern:2001dg}, to which we refer the reader for a more detailed discussion. As was done there, we treat the coupling associated to the external photons differently from the ones associated to virtual photons and gluons. In particular, we renormalise the external photon fields and their couplings to the internal fermion lines in the on-shell (OS) scheme, which simply amounts to the following modification of \cref{eq:hel_1},
\begin{equation}
    \mathcal{M}_{{\rm bare}, \lambdavec} = (8 \alpha_b^2)\, A_{ \lambdavec}(\alpha_b,\alpha_{s,b}) 
    \quad \rightarrow \quad 
    \mathcal{M}_{{\rm bare}, \lambdavec} = (8 \alpha_{\text{OS}}^2)\, A_{ \lambdavec}(\alpha_b,\alpha_{s,b}) \,,
\end{equation}
where $\alpha_{\text{OS}} \simeq 1/137$ is the on-shell QED coupling constant. In what follows, for simplicity and better readability, we will drop the subscript related to the scheme choice.

We perform UV renormalization of the remaining couplings in $\overline{\rm MS}$: 
\begin{equation}\label{eq:asren}
\begin{aligned}
    S_{\epsilon}\, \mu_0^{2\ep}\,\alpha_{s,b} &= \mu^{2\ep}\, \alpha_s(\mu)\, Z_{\text{QCD}}[{\alpha_s(\mu)}]\, , \\
    S_{\epsilon}\, \mu_0^{2\ep}\,\alpha_{b} &= \nu^{2\ep}\, \alpha(\nu)\, Z_{\text{QED}}[{\alpha(\nu)}] \,,
\end{aligned}
\end{equation}
with $\mu$ the QCD renormalisation scale and $\nu$ the QED one.
Above we used $S_{\epsilon} = (4\pi)^{\epsilon}e^{-\gamma_E \epsilon}$, the renormalisation factors
\begin{equation}\label{eq:renorm}
\begin{aligned}
Z_{\text{QCD}}[\alpha_s] = 1 - \frac{\beta^{\text{QCD}}_0}{\ep}\left(\frac{\alpha_s}{2\pi}\right) + O(\alpha_s^2) \,, \quad\quad
Z_{\text{QED}}[\alpha] = 1 - \frac{\beta^{\text{QED}}_0}{\ep}\left(\frac{\alpha}{2\pi}\right) + O(\alpha^2) \,,
\end{aligned}
\end{equation}
and the one-loop $\beta$-functions
\begin{equation}
\beta^{\text{QCD}}_0 = \frac{11}{6}C_A - \frac{2}{3}T_F n_f \,, \quad\quad
\beta^{\text{QED}}_0 = - \frac{2}{3} (n_l + N_c\, \tilde{n}_q^{(2)})\, .
\end{equation}
Note that at this perturbative order the renormalisation of one coupling does not enter the one of the other, as reflected by the equations above. Starting at the next order one would have to consider mixing in the renormalisation of $\alpha$ and $\alpha_s$.

Plugging \cref{eq:asren} into \cref{eq:full_bare_amplitude} and expanding to the appropriate perturbative orders in $\alpha$ and $\alpha_s$, we obtain a perturbative series involving renormalised amplitudes which we define as follows:\footnote{We use this definition of renormalised amplitudes in order to make the dependence on the QED and QCD renormalisation scales explicit. Note however that this definition of renormalised amplitudes differs from the ones used in \cite{Caola:2020dfu,Caola:2021rqz,Caola:2021izf,Bargiela:2021wuy,Caola:2022dfa,Bargiela:2022lxz}. }
\begin{equation}\label{eq:full_amplitude}
\begin{split}
    {f}_{\text{QCD+QED,ren},\vec\lambda} &= \widetilde{\mathcal{N}}^{(1,1)} \, {{f}}^{(1,1)}_{\text{ren},\vec\lambda} +
    \widetilde{\mathcal{N}}^{(2,1)} \, {{f}}^{(2,1)}_{\text{ren},\vec\lambda} + 
    \sum_{j=1}^4\widetilde{\mathcal{N}}^{(3,j)} \, {{f}}^{(3,j)}_{\text{ren},\vec\lambda}
    \\
    & + \left( \widetilde{\mathcal{N}}^{(3)}_{\text{QCD}} \, L(\mu) + \widetilde{\mathcal{N}}^{(3)}_{\text{QED}}\, L(\nu) \right) f^{(2,1)}_{\text{ren},\vec\lambda} \,,
\end{split}
\end{equation}
where we have introduced the shorthand notation
\begin{equation}
    L(\mu)=\log\left(\frac{\mu^2}{s}\right) \, ,
\end{equation} 
and defined 
\begin{equation}
    \widetilde{\mathcal{N}}^{(3)}_{\text{QCD}} = \left( \frac{\alpha_s}{\pi} \right)^2 \, N_c\, C_F \, \tilde{n}_q^{(4)} \, \frac{\beta^{\text{QCD}}_0}{2} \,,
    \quad\quad
    \widetilde{\mathcal{N}}^{(3)}_{\text{QED}} = \left(\frac{\alpha}{\pi}\right)^2 \left(n_l + N_c\, \tilde{n}_q^{(6)} \right) \, \frac{\beta_0^{\text{QED}}}{2} \,, 
\end{equation}
The second line of \cref{eq:full_amplitude} makes the dependence of the ($\mathcal{O}(\ep^0)$ part of the) renormalised amplitude on the renormalisation scales explicit, and its two terms arise directly from the renormalisation of the two-loop $\widetilde{\mathcal{N}}^{(2,1)} \, {{f}}^{(2,1)}_{\vec\lambda}$ term.
Indeed one can check that $\widetilde{\mathcal{N}}^{(3)}_{\text{QCD}} + \widetilde{\mathcal{N}}^{(3)}_{\text{QED}} = \widetilde{\mathcal{N}}^{(2,1)}$, implying that when one sets $\mu=\nu$ the full renormalised amplitude can be computed as the abelianisation of the renormalised pure QCD amplitude. Nevertheless, here we preferred to keep the renormalisation scales of the two couplings separate. 

The renormalised ${{f}}^{(L,i)}_{\text{ren},\vec\lambda}$ above have been defined so to be free of poles in $\ep$ as well as independent of the renormalisation scales $\mu$ and $\nu$ at $\mathcal{O}(\epsilon^0)$.
As no further IR regularization is required, each renormalised amplitude $f^{(L)}_{\text{ren},\vec\lambda}$ is the $L$-loop physical finite remainder.


Finally, let us present our results for the one-loop, two-loop, and three-loop helicity amplitude finite remainders.
We observe the final formulae are remarkably compact despite lengthy intermediate expressions.
The three-loop result is linearly decomposed into products of 126 HPLs to weight 6 and their coefficients, which are proportional to $\frac{x^n}{(1-x)^m}$ for $-2 \leq n \leq 2$ and $-2 \leq m \leq -1$.
When decomposed to minimal set of Lyndon words, this linear basis of 126 HPLs can be expressed in terms of products of 23 HPLs~\cite{Duhr:2019tlz}.
We note that there is a weight drop of 3 for all-plus and 2 for single-minus comparing to the full weight 6 double-minus MHV configuration.
Due to the compactness of the all-plus result, we can report it here:
\allowdisplaybreaks
\begin{align}
{f}_{\text{ren},++++}^{(1,1)} =& 1 \,, \nonumber \\
{f}_{\text{ren},++++}^{(2,1)} =& -\frac{3}{2} \,, \nonumber \\
{f}_{\text{ren},++++}^{(3,1)} =& 
\left(-\frac{x^2}{8}-\frac{1}{8x^2}+\frac{x}{8}+\frac{1}{8 x}-\frac{1}{8}\right) L_1^2
+\left(-\frac{i \pi }{4 x^2}-\frac{x}{4}+\frac{(i \pi - 1)}{4 x}+\frac{(2-i \pi )}{8}\right)L_1 \nonumber \\
& + \left(\frac{x^2}{8} - \frac{x}{8} + \frac{1}{16}\right) L_0 \, L_1 -\frac{\pi ^2 x^2}{16} +\frac{\pi ^2 x}{16}+\frac{i \pi }{8 (x-1)}-\frac{i \pi }{8 x} +\frac{\left(39+4 i \pi -\pi ^2\right)}{32} \nonumber \\
& + \{(x) \leftrightarrow (1-x)\} \,, \nonumber \\
{f}_{\text{ren},++++}^{(3,2)} =& 
\left(-\frac{11 x^2}{48}-\frac{11}{48 x^2}+\frac{11 x}{48}+\frac{11}{48 x}\right) L_1^2
+\left(-\frac{11 i \pi }{24 x^2}-\frac{11 x}{24}+\frac{11 (i \pi - 1)}{24 x}+\frac{11}{12}\right) L_1 \nonumber \\
& + \left(\frac{11 x^2}{48} - \frac{11 x}{48}\right) L_0 \, L_1 -\frac{11 \pi ^2 x^2}{96}+\frac{11 \pi ^2 x}{96}+\frac{11 i \pi }{48 (x-1)}-\frac{11 i \pi }{48 x} -\frac{175}{48}  \nonumber \\
& + \{(x) \leftrightarrow (1-x)\} \,, \nonumber \\
{f}_{\text{ren},++++}^{(3,3)} =& 
\left(\frac{x^2}{12}+\frac{1}{12 x^2}-\frac{x}{12}-\frac{1}{12 x}\right) L_1^2
+\left(\frac{i \pi }{6 x^2}+\frac{x}{6}-\frac{(i \pi - 1)}{6 x}-\frac{1}{3}\right) L_1 \nonumber \\
& + \left(-\frac{x^2}{12}+\frac{x}{12}\right) L_0 \, L_1 +\frac{\pi ^2 x^2}{24}-\frac{\pi ^2 x}{24}-\frac{i \pi }{12 (x-1)}+\frac{i \pi }{12 x} +\frac{7}{6} \nonumber \\
&
+ \{(x) \leftrightarrow (1-x)\} \,, \nonumber \\
{f}_{\text{ren},++++}^{(3,4)} =& 
\left(\frac{3 x^2}{4}+\frac{3}{4 x^2}-\frac{3 x}{4}-\frac{3}{4 x}+\frac{1}{2}\right) L_1^2
+\left(\frac{3 i \pi }{2 x^2}+2 x-\frac{ (3 i \pi - 4)}{2 x}+\frac{(i \pi - 4)}{2}\right) L_1 \nonumber \\
& + \left(-\frac{3x^2}{4} + \frac{3x}{4} -\frac{1}{4} \right) L_0 \, L_1 +\frac{3 \pi ^2 x^2}{8}-\frac{3 \pi ^2 x}{8}-\frac{i \pi }{x-1}+\frac{i \pi }{x} \nonumber \\
&+\frac{\left(24 \zeta_3 + 2-8 i \pi +\pi ^2\right)}{8}
+ \{(x) \leftrightarrow (1-x)\} \,,
\end{align}
where we defined $L_0=\log(x)$ and $L_1=\log(1-x)$. 

\begin{figure}[t]
\centering
\begin{subfigure}[b]{0.49\textwidth}
\includegraphics[width=\textwidth]{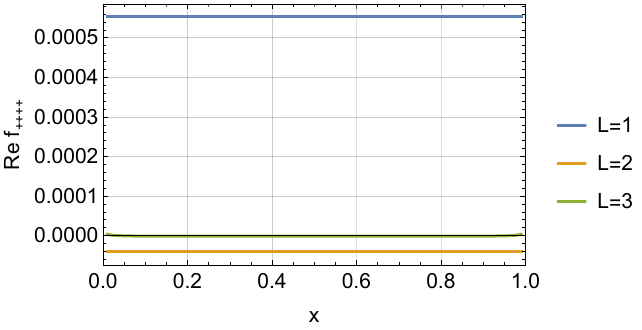}
\end{subfigure}
\begin{subfigure}[b]{0.5\textwidth}
\includegraphics[width=\textwidth]{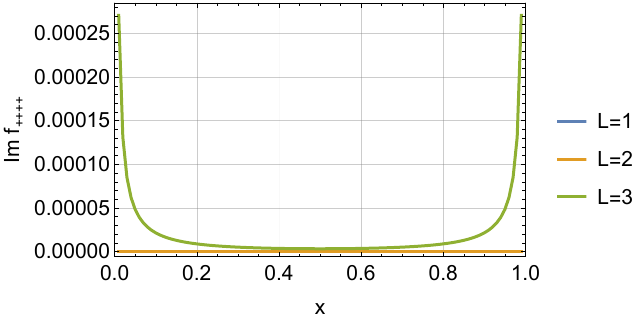}
\end{subfigure}
\begin{subfigure}[b]{0.49\textwidth}
\includegraphics[width=\textwidth]{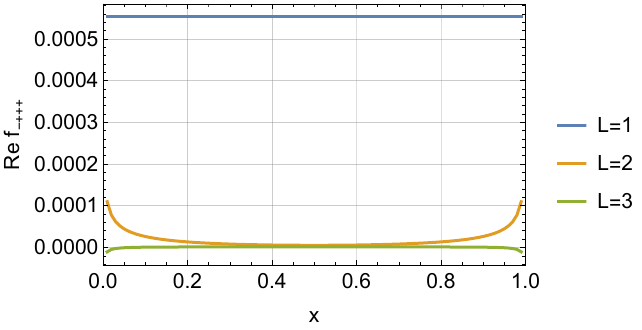}
\end{subfigure}
\begin{subfigure}[b]{0.5\textwidth}
\includegraphics[width=\textwidth]{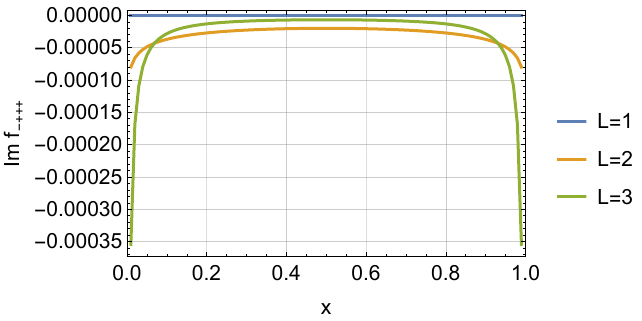}
\end{subfigure}
\begin{subfigure}[b]{0.49\textwidth}
\includegraphics[width=\textwidth]{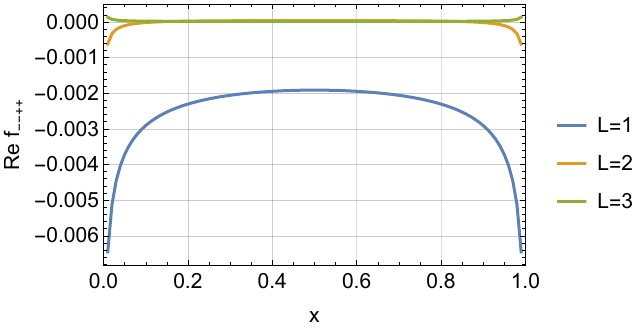}
\end{subfigure}
\begin{subfigure}[b]{0.5\textwidth}
\includegraphics[width=\textwidth]{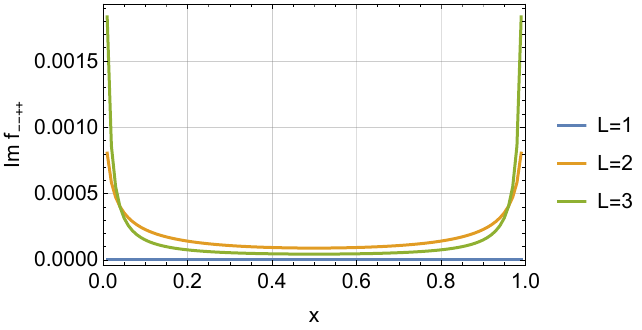}
\end{subfigure}
\caption{QCD finite remainders with coupling constants,
$8\alpha^2\, \left({\alpha_s}/{\pi}\right)^{L-1}\, f^{(L)}_{\text{ren},\lambdavec}$, as function of $x=-t/s$ for specific values of $\alpha=1/137$, $\alpha_s=0.118$ and with $n_q=5$ quark flavours.}
\label{fig:numhel}
\end{figure}

As our final formulae can be evaluated numerically in a fast and efficient way, we plot the QCD finite remainders including all coupling constants, $8\alpha^2\, \left({\alpha_s}/{\pi}\right)^{L-1}\, f^{(L)}_{\text{ren},\lambdavec}$, in \cref{fig:numhel} for the three independent helicity configurations at the different orders in the $\alpha_s$ expansion. 
As for the coupling parameters, we use $\alpha=1/137$ and $\alpha_s=0.118$ with $n_q=5$ quarks flavours in the fermion loops, and with the corresponding $\tilde{n}_q^{(i)}$ values. We set the QCD renormalisation scale to $\mu^2=s$. 
We note that while for the real part the contribution from the helicity amplitudes get smaller at higher loop order, their contribution gets enhanced for the imaginary part. In particular, we remark that the one-loop amplitude has no imaginary contribution as this is a loop-induced process. The magnitude of the contributions tends to be always enhanced in the (anti-)collinear regions at the end points $x=0$ and $x=1$.

In the ancillary material that accompany this manuscript, we provide our analytic results in a computer-readable format for the finite reminders of the 5 helicity amplitudes $f_{++++}$, $f_{-+++}$, $f_{--++}$, $f_{+-+-}$, and $f_{+--+}$ to transcendental weight 6 at one-, two-, and three-loop order written both in terms of HPLs, and alternatively in terms of a minimal set of multiple polylogarithms defined as
\begin{equation}
\label{eq:Linm}
\Li_{m_1,...,m_k}(x_1,...,x_k) = \sum_{i_1>...>i_k>0} \frac{x_1^{i_1}}{i_1^{m_1}} \, ... \, \frac{x_k^{i_k}}{i_k^{m_k}} \,,
\end{equation}
following ref.~\cite{Vollinga:2004sn}.
This basis of 23 transcendental functions consists of 2 logarithms, $\log(x)$, $\log(1-x)$, 12 classical polylogarithms, $\Li_2$ of $x$, $\Li_3$ of $x$ and $(1-x)$, and $\Li_4$, $\Li_5$, $\Li_6$ of $x, (1-x)$ and $-x/(1-x)$, as well as 9 multiple polylogarithms, $\Li_{3,2}(1,x)$, $\Li_{3,2}(1-x,1)$, $\Li_{3,2}(x,1)$, $\Li_{3,3}(1-x,1)$, $\Li_{3,3}(x,1)$, $\Li_{3,3}\left(\frac{-x}{1-x},1\right)$, $\Li_{4,2}(1-x,1)$, $\Li_{4,2}(x,1)$, $\Li_{2,2,2}(x,1,1)$.

\paragraph{Checks}
We analytically checked our results at one and two loops against the literature~\cite{Bern:2001dg} and found perfect agreement. We were also able to cross-check analytically, at the three-loop bare level, the abelian-type contribution ${{f}}^{(3,1)}_{\vec\lambda}$ against the corresponding contribution from the $gg\rightarrow \gamma \gamma$ computation in ref.~\cite{Bargiela:2021wuy} and again found agreement. In addition, after accounting for the different colour structures, we also succeeded in the cross-check of the two-fermion type contributions, which is the combined sum of ${{f}}^{(3,3)}_{\vec\lambda}$ and ${{f}}^{(3,4)}_{\vec\lambda}$ at the bare level (see \cref{fig:3loop2,fig:3loop3}), against the corresponding one in ref.~\cite{Bargiela:2021wuy} and found full agreement. The new pieces in the calculation is the non-abelian set ${{f}}^{(3,2)}_{\vec\lambda}$ and the separation of the sets ${{f}}^{(3,3)}_{\vec\lambda}$ and ${{f}}^{(3,4)}_{\vec\lambda}$.
\section{Differential cross sections}
\label{subsec:pheno}

In this section, we briefly discuss the phenomenological application of the scattering amplitudes. Light-by-Light scattering has been experimentally observed at the LHC by both the ATLAS \cite{ATLAS:2017fur,ATLAS:2019azn,ATLAS:2020hii} and the CMS \cite{CMS:2018erd,CMS:2024tfd} collaborations within Ultra-Peripheral Collisions (UPC) of two heavy ions (PbPb). These events occurred at a nucleon-nucleon centre-of-mass energy of $\sqrt{s_{NN}}=5.02$ TeV. Due to detector constraints, one has, in the case of ATLAS, to apply a few kinematic cuts as follows
\begin{equation}
    m_{\gamma\gamma}>5\text{ GeV}, \qquad |p_{t,\gamma}|>2.5\text{ GeV}, \qquad |\eta_{\gamma}|<2.37,
    \label{eq:cuts}
\end{equation}
where $m_{\gamma \gamma}$ is the invariant mass of the di-photon pair, $|p_{t,\gamma}|$ is the photon transverse momentum and $|\eta_{\gamma}|$ is the photon (pseudo-)rapidity.

The phase-space integrated cross section without cuts can be expressed as follows
\begin{equation}
    \sigma_{\gamma \gamma} = \int_{0}^1 {\rm d}\tau \int^{-\frac{1}{2}\log{\tau}}_{\frac{1}{2}\log{\tau}} {\rm d}y\; \mathcal{L}^{(AB)}{(\tau,y)}\, \frac{1}{\tau^2 s_{NN}^2} \frac{1}{16 \pi} \int_{-\tau s_{NN}}^0 {\rm d}t\; \overline{|\mathcal{M}(\tau,y,t)|^2} \,,
    \label{eq:crossform}
\end{equation}
where $\mathcal{L}^{(AB)}{(\tau,y)}$ is the luminosity factor of the two photons from the heavy ions $A$ and $B$. 
As the process under consideration is an exclusive one and does not involve any additional parton in the final state, we can associate the integration variables $\tau$ and $y$ with the invariant di-photon mass, $m_{\gamma \gamma} = \sqrt{\tau s_{NN}}$, and the combined rapidity of the di-photon system in the lab frame, $y_{\gamma \gamma}=y$. 

The luminosity factor can be computed using dedicated software such as {\textsc{gamma}-UPC} \cite{Shao:2022cly}, where we employ the charge form factor (ChFF) method provided in that package. Above, $\overline{|\mathcal{M}|^2}$ indicates the sum over all helicity states, and includes the average over the initial states and the symmetry factor for identical final-state particles as
\begin{align}
    \overline{|\mathcal{M}|^2} = \frac{1}{2!} \frac{1}{2^2} \sum_{\rm \vec\lambda} |\mathcal{M}_{\vec\lambda}|^2.
\end{align}

In the following, taking into account the kinematic cuts defined in eq.~(\ref{eq:cuts}), we focus on the two differential distributions, $d\sigma_{\gamma \gamma}/dm_{\gamma \gamma}$ and $d\sigma_{\gamma \gamma}/dy_{\gamma \gamma}$. We compute these observables up to NNLO in both QCD and QED and also include their mixed QCD-QED contribution. Due to the exclusive nature of the process, we can consider the NNLO corrections at the level of the amplitude rather than at cross-section level itself.

Using the results from previous sections, we can express the full renormalised helicity amplitudes up to NNLO as follows
\begin{align}
    \mathcal{M}^{\text{LO}}_{\vec\lambda} &= (8\alpha^2)\left[ \widetilde{\mathcal{N}}^{(1,1)} \, {{f}}^{(1,1)}_{\text{ren},\vec\lambda}
    \right] S_{\vec\lambda}
    \,,
    \\
    \mathcal{M}^{\text{NLO}}_{\vec\lambda} &= (8\alpha^2)\left[ \widetilde{\mathcal{N}}^{(1,1)} \, {{f}}^{(1,1)}_{\text{ren},\vec\lambda} +
    \widetilde{\mathcal{N}}^{(2,1)} \, {{f}}^{(2,1)}_{\text{ren},\vec\lambda} 
    \right] S_{\vec\lambda}
    \,,
    \\
    \begin{split}
    \mathcal{M}^{\text{NNLO}}_{\vec\lambda} &= (8\alpha^2)\Biggl[
    \widetilde{\mathcal{N}}^{(1,1)} \, {{f}}^{(1,1)}_{\text{ren},\vec\lambda} +
    \widetilde{\mathcal{N}}^{(2,1)} \, {{f}}^{(2,1)}_{\text{ren},\vec\lambda} + 
    \sum_{j=1}^4\widetilde{\mathcal{N}}^{(3,j)} \, {{f}}^{(3,j)}_{\text{ren},\vec\lambda} 
    \\
    &  + \left( 
    \widetilde{\mathcal{N}}^{(3)}_{\text{QCD}} \, L(\mu) +
    \widetilde{\mathcal{N}}^{(3)}_{\text{QED}}\, L(\nu)
    \right) f^{(2,1)}_{\text{ren},\vec\lambda} 
    \Biggl] S_{\vec\lambda}
    \, ,
    \end{split}
\end{align}
where it is understood that from the two-loop order on, the helicity amplitudes have a dependence on the QCD renormalisation scale $\mu$. While this dependence is implicit in the running of the QCD coupling $\alpha_s$, the explicit dependence appears only at the three-loop level. Above, when evaluating $|\mathcal{M}_{\vec\lambda}|^2$ one can ignore the spinor factors $S_{\vec\lambda}$ since they are defined so that $|S_{\vec\lambda}|^2 = 1$.

In order to evaluate the strong coupling constant at the different scales, we make use of the \texttt{RunDec} package \cite{Herren:2017osy}. While, for the NLO QCD contribution, we evolve the coupling up to one-loop level, we compute $\alpha_s$, for the NNLO QCD contributions, up to two-loop order. In order to assess the uncertainty related to the QCD corrections, we vary the renormalisation scale $\mu$ around its central value $\mu_{\text{cent.}} = \sqrt{s}/2$ by a factor of two, $\mu = [ \sqrt{s}/4, \sqrt{s} ]$.

In the case of the QED coupling, irrespective of the correction order, we make use of its value at the on-shell point as
\begin{equation}
    \alpha = \frac{1}{137}\,.
\end{equation}
This is despite the fact that the three-loop QED amplitude actually exhibits an explicit dependence on the QED renormalisation scale $\nu$. Having performed the QED renormalisation within the $\overline{\rm MS}$ scheme, the coupling should in principle evolve with varying $\nu$. However, we believe that the impact of evolving the QED coupling is negligible compared to the QCD case, hence set throughout the computation $\nu = \sqrt{s}/2$.

In the amplitude we consider electron, muon and tau leptons inside the closed fermion loops and hence set $n_l=3$. The two-particle threshold for the tau lepton lies below the kinematic cut-off for the di-photon invariant mass in eq.~(\ref{eq:cuts}), and is therefore considered to be massless in the computation. In the case of quarks, we set $n_q=4$ below the two-particle threshold for bottom quarks, $\sqrt{s} < 2m_b = 9.5\text{ GeV}$, and set $n_q=5$ above. This corresponds for the fractional charge factors,
\begin{equation}
\begin{split}
    \tilde{n}_q^{(2)}=&\begin{cases} \frac{10}{9} & n_q=4\,, \\ \frac{11}{9} & n_q=5\,, \end{cases} \qquad \tilde{n}_q^{(4)}=\begin{cases} \frac{34}{81} & n_q=4\,, \\ \frac{35}{81} & n_q=5\,, \end{cases}
    \\
    \tilde{n}_q^{(6)}=&\begin{cases} \frac{130}{729} & n_q=4\,, \\ \frac{131}{729} & n_q=5\,, \end{cases} \qquad \tilde{n}_q^{(8)}=\begin{cases} \frac{514}{6561} & n_q=4\,, \\ \frac{515}{6561} & n_q=5\,. \end{cases}
\end{split}
\end{equation}
Having discussed the phenomenological parameters and the computational setup, we are now ready to discuss the cross-section results. In order to assess the size of the QCD and QED corrections, we define the $K$-factor as
\begin{equation}
    K^{\text{NLO}} = \frac{d\sigma^{\text{NLO}}/dx}{d\sigma^{\text{LO}}/dx}\,, \qquad\qquad K^{\text{NNLO}} = \frac{d\sigma^{\text{NNLO}}/dx}{d\sigma^{\text{LO}}/dx}\,,
\end{equation}
where $x$ represents the corresponding distribution parameter. As for cross-checks of our setup, we have evaluated the LO cross section with the MadGraph event generator \cite{Alwall:2014hca}, that evaluates at most one-loop diagrams, and found good agreement with our LO results.

\begin{figure}
\begin{center}
\subfloat[]{\includegraphics[width=0.49\textwidth]{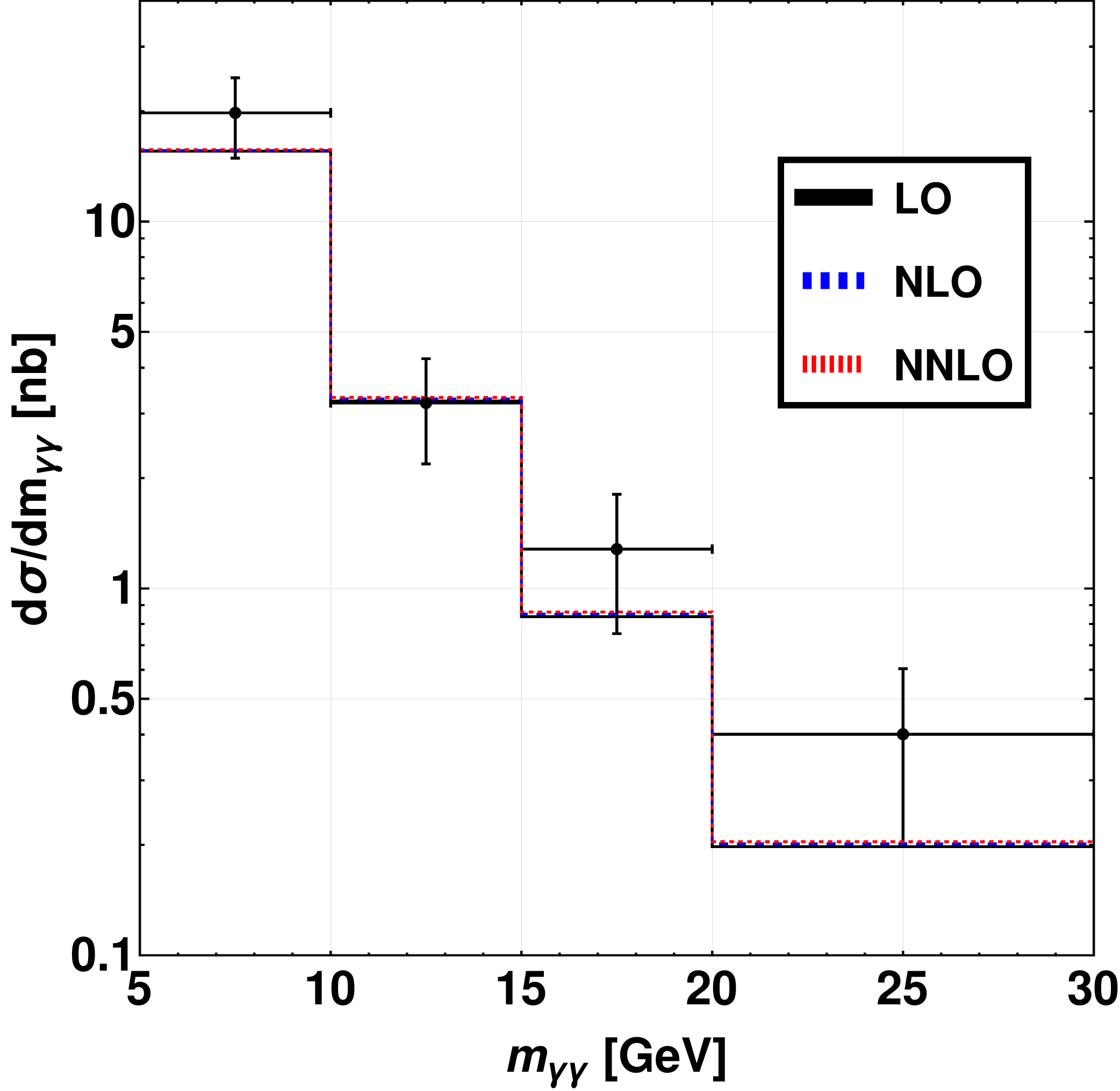}
\label{fig:dsigdmggbin}}
\subfloat[]{\includegraphics[width=0.49\textwidth]{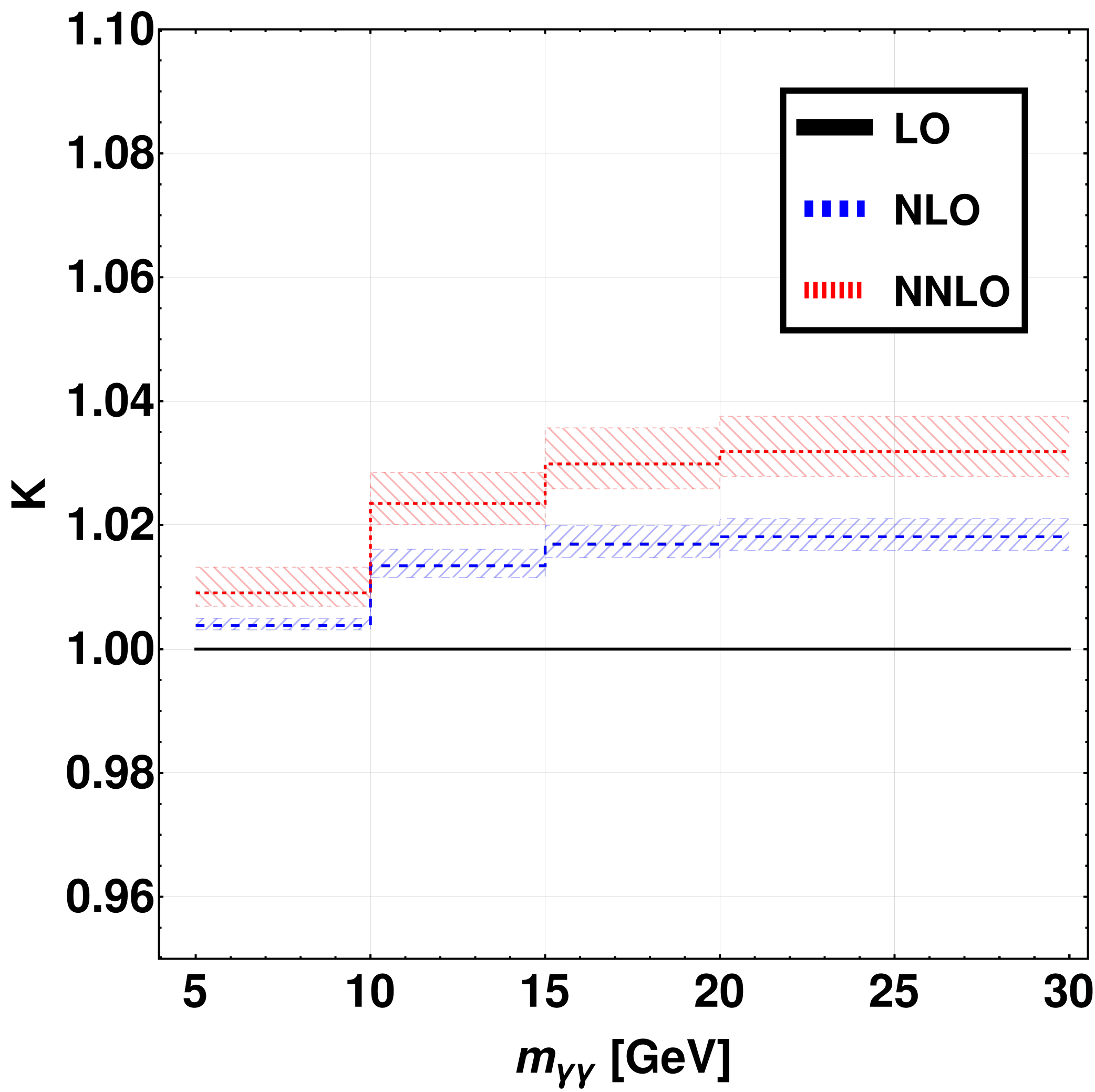}
\label{fig:Kmgg}}
\end{center}
\vspace{-0.5cm}
\caption{Differential distribution in (a) $d\sigma/dm_{\gamma \gamma}$ and (b) corresponding $K$-factor at LO, NLO and NNLO with bin sizes corresponding to experimental ATLAS data \cite{ATLAS:2020hii}.}
\label{fig:dsigdmgg}
\end{figure}

In \cref{fig:dsigdmgg}, we plot the differential distributions in the invariant di-photon mass $m_{\gamma \gamma}$ with the corresponding $K$-factor at LO, NLO and NNLO. In order to compare with the experimental ATLAS data \cite{ATLAS:2020hii}, we employ the same bin size in the $m_{\gamma \gamma}$ distribution. As can been seen in \cref{fig:dsigdmggbin}, we find that, while the LO cross section agrees with the experimental data for all data bins except the last one, where it lies marginally below, both the NLO and the NNLO predictions are consistent with all data bins.

As can be observed in \cref{fig:Kmgg}, while the size of the NLO contributions is on the order of $\mathcal{O}{\left(0.5\%\right)}$ at low invariant masses, it can increase up to $\mathcal{O}{\left(2\%\right)}$ at the last bin. In contrast to this, the size of the NNLO contributions is always larger, and can reach corrections of the size of $\mathcal{O}{\left(1\%\right)}$ already at the first bin up to $\mathcal{O}{\left(3.5\%\right)}$ at the last bin. We also note that the LO, NLO and NNLO results do not overlap within their uncertainties, and that the relative uncertainty of the NNLO result turns out to be slightly larger than the NLO result.

An explanation for the relative large size of the NNLO correction lies in the new topology represented by the gauge-invariant set $\mathcal{N}^{(3,4)}$ (see \cref{fig:3loop2} and \cref{eq:three-loop-colour-def-QCD,eq:three-loop-colour-def-MIXED}) which appears for the first time at the three-loop level. A simple check at the amplitudes $\widetilde{\mathcal{N}}^{(3,j)} \, {{f}}^{(3,j)}_{\text{ren},\vec\lambda}$ reveals that the new topology, $j=4$, is significantly enhanced compared to the other gauge-invariant contributions.

\begin{figure}
\begin{center}
\subfloat[]{\includegraphics[width=0.485\textwidth]{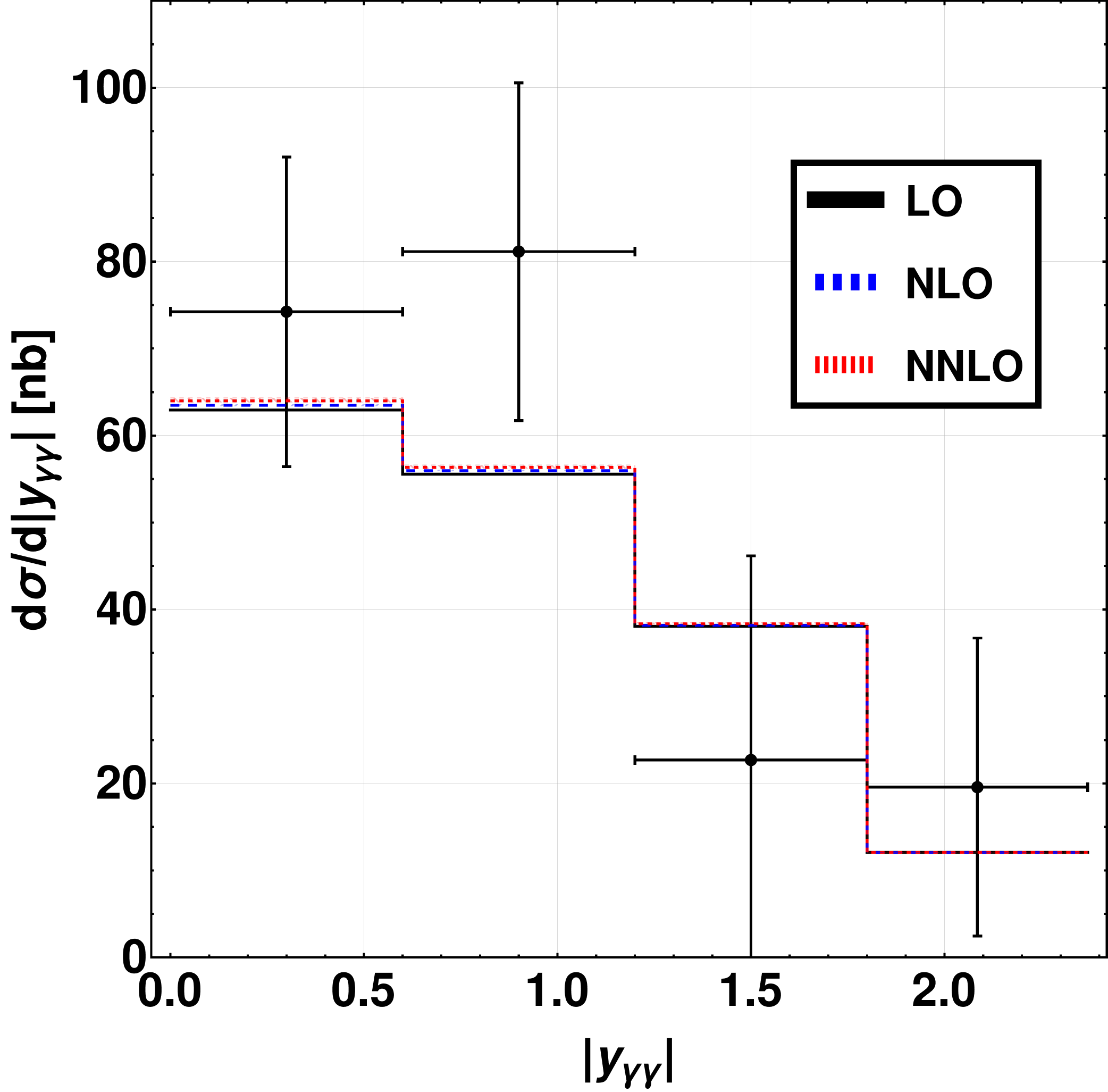}
\label{fig:dsigdyggbin}}
\subfloat[]{\includegraphics[width=0.49\textwidth]{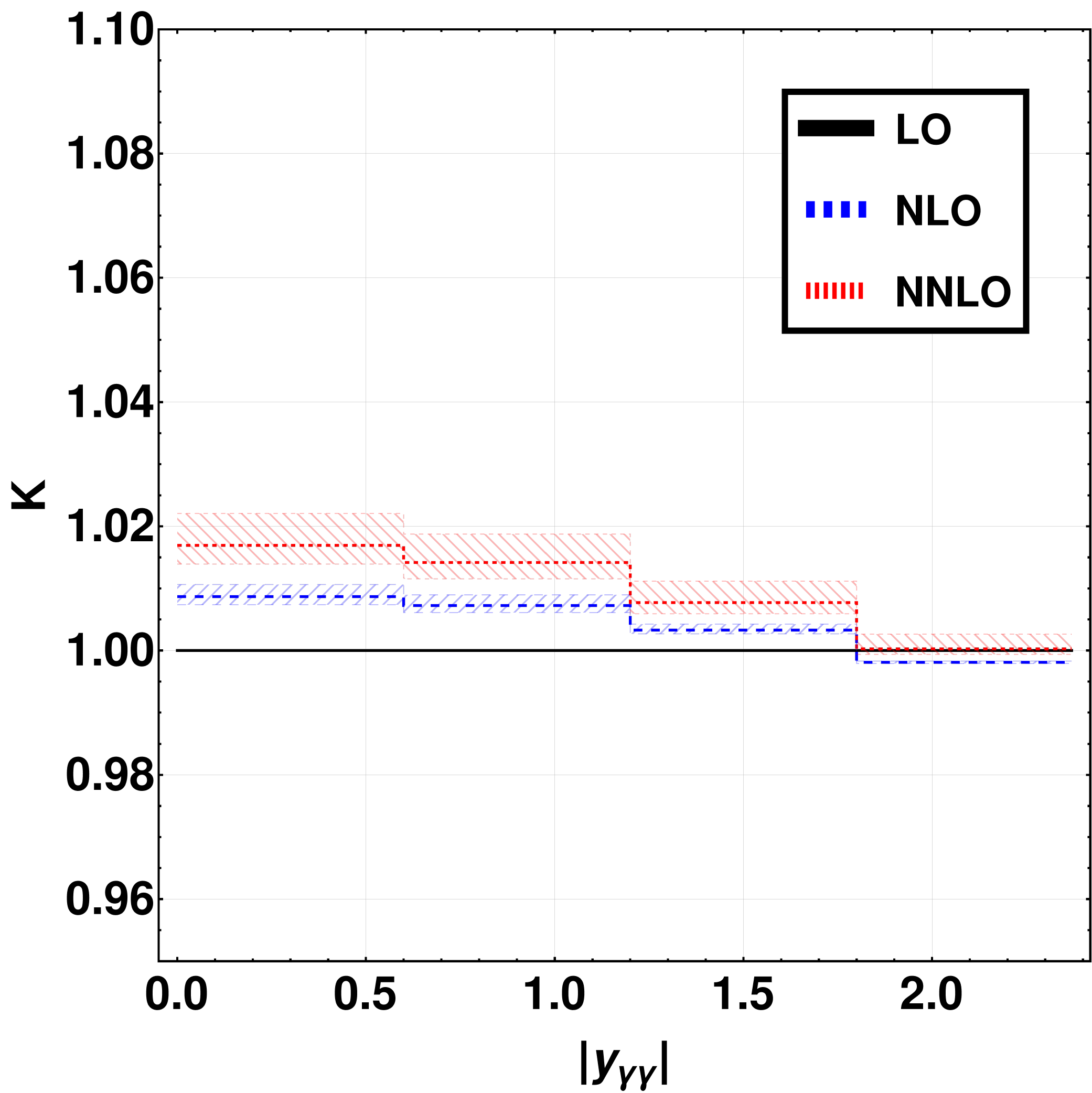}
\label{fig:Kygg}}
\end{center}
\vspace{-0.5cm}
\caption{Differential distributions in (a) $d\sigma/d|y_{\gamma \gamma}|$ and (b) corresponding $K$-factor at LO, NLO and NNLO with bin sizes corresponding to experimental ATLAS data \cite{ATLAS:2020hii}.}
\label{fig:dsigdygg}
\end{figure}

In \cref{fig:dsigdygg}, we display the differential distributions in the rapidity of the di-photon pair $|y_{\gamma \gamma}|$ and show also its $K$-factor. We observe that all LO, NLO and NNLO results agree with all data bins except the second bin. It should be noted that only a few events were measured, hence the experimental data exhibit quite large uncertainties. This is expected to be improved in future measurements.

Similarly as before, we find that both the QCD and QED corrections are relatively large in the central rapidity region compared to the strong forward or backward regions. While the NLO result introduces corrections of around $\mathcal{O}{\left(1\%\right)}$ in the central region and up to $\mathcal{O}{\left(0.1\%\right)}$ at the forward and backward region, the NNLO corrections are roughly twice as large with $\mathcal{O}{\left(2\%\right)}$ correction in the central regions but again up to $\mathcal{O}{\left(0.1\%\right)}$ level at the boundaries.

Regarding the total phase-space integrated cross section, our NNLO prediction is
\begin{equation}
    \sigma^{\text{NNLO}} = 102.10^{+0.44}_{-0.25} \, \text{nb}\,,
\end{equation}
which is consistent with the experimental data from the ATLAS experiment, $\sigma_{\text{ATLAS}} = 120 \pm 22 \, \text{nb}$.

At this stage, we would like to make a small comment regarding the comparison to the massive case which was considered up to NLO accuracy in ref.~\cite{AH:2023kor}. We note that, by considering the observable $d\sigma/dm_{\gamma \gamma}$, already the LO result in the massive case \cite{AH:2023kor} does not agree with the first bin which corresponds to the low invariant mass distribution. Relevant in the calculation are the massive charm and bottom quarks and the massive tau lepton.

It is clear that amplitudes are suppressed in the massive case compared to the massless one. Regarding the size of the corrections in the massive case, the NLO contribution for that first bin were found to be on the order of $\mathcal{O}{\left(7.5\%\right)}$ in ref.~\cite{AH:2023kor}. Observing that in our massless computation the NNLO contribution is almost as large as the NLO contribution, we anticipate similarly large corrections also in the massive case, which we leave for future work. 

\section{Conclusions}
\label{sec:concl}

In this work, we computed all independent helicity amplitudes for the $\gamma\gamma \to \gamma\gamma$ process in three-loop massless QCD and QED, as well as the corresponding NNLO differential cross section in the invariant mass and rapidity distribution of the di-photon system.
This high-precision prediction is relevant for phenomenology of the UPC of heavy ions at the LHC.

In order to tame the amplitude computational complexity, we employed modern methods in tensor and integral reduction.
In particular, we used the ’t Hooft-Veltman tensor decomposition which minimizes the number of form factors that contribute to the helicity amplitudes.
In addition, we exploited syzygies and finite-field arithmetics in the IBP reduction to MIs.
Despite the presence of large intermediate expressions, our final formulae for the finite part of helicity amplitudes are remarkably compact.
They can be expressed in terms of 23 HPLs with letters $\{0,1\}$ of transcendental weight up to 6, and can be evaluated numerically in a fast and stable manner.
We provide our analytic results in the ancillary files that accompany this publication.

In order to promote our amplitudes to hadronic cross sections, we convoluted their modulus squared with the luminosity of the two photons emitted by the heavy ions.
We computed the differential distributions both in the invariant mass and the rapidity of the di-photon system up to NNLO in both QCD and QED by including also their mixed contribution.
By running the strong and electromagnetic couplings in different schemes, we estimate the missing higher-order uncertainty to our NNLO predictions.
We performed a detailed analysis of the $K$-factors of our observables and
note that, as can be seen from \cref{fig:dsigdmgg,fig:dsigdygg}, the NNLO corrections are as large as the NLO one for every considered value of the kinematic phase space.
In particular, we observe that all the considered higher-order predictions exceed their lower-order uncertainty estimates.
Comparing our differential cross-section predictions with the experimental data, we find that our higher-order results are largely in agreement with the ATLAS data.

Phenomenologically, a natural future direction is to include contributions from massive quark and lepton loops in the differential cross sections.
Indeed, at low invariant mass of the di-photon system, the inclusion of massive charm and bottom quark and massive tau lepton corrections may be important around their production thresholds.
At higher invariant mass, massive top quark loops could be considered, although their contribution is expected to remain minor.
In a more mathematical direction, the massless three-loop amplitude may serve as a testing ground for identifying more direct routes to the simplicity of the final result. 
We look forward to expand on these investigations in the future.

\section*{Acknowledgements}
We thank Ajjath AH, Ekta Chaubey and Hua-Sheng Shao for interesting exchanges on their light-by-light computation in ref.~\cite{AH:2023kor}. We thank Lukas Simon and Hua-Sheng Shao for help regarding the {\textsc{gamma}-UPC} \cite{Shao:2022cly} and the MadGraph \cite{Alwall:2014hca} packages. We are grateful to Andreas von Manteuffel for allowing us to use the reduction tables he previously computed with \texttt{Finred} for our earlier projects. We are also grateful to Taushif Ahmed, Charalampos Anastasiou, \'Emilien Chapon, Andreas von Manteuffel, Vajravelu Ravindran and Raoul Röntsch for useful discussions.
The research of PB was supported by the Swiss National Science Foundation (SNF) under contract 200020-204200, by the European Research Council (ERC) under the European Union's Horizon 2020 research and innovation programme grant agreement 101019620 (ERC Advanced Grant TOPUP), and the ERC under the European Union’s Horizon Europe research and innovation program grant agreement 101163627 (ERC Starting Grant “AmpBoot”). AC has been supported by the Italian Ministry of
Universities and Research through Grant No. PRIN 2022BCXSW9. 

\bibliographystyle{JHEP}
\bibliography{references}

\end{document}